%% file: main.tex
\documentclass[sigconf, nonacm]{acmart}

\usepackage{xspace}
\usepackage{soul}
\usepackage{xcolor}
\usepackage{enumitem} 
\usepackage{marginnote}
 \usepackage{multirow}
 \usepackage{graphicx}
 \usepackage{booktabs}
 \usepackage[section]{placeins}
 \usepackage{xurl}
\newcommand{\tinyskip}{\vspace{3pt}}
\newcommand{\mypar}[1]{\tinyskip\noindent\textbf{#1.}\xspace}

\newcommand{\F}{\mbox{Fig.\hspace{0.25em}}}
\newcommand{\KR}{MM}

\newenvironment{myitemize}{%
\begin{itemize}[leftmargin=1em, itemsep=.1em, parsep=.1em, topsep=.1em,
    partopsep=.1em]}
{\end{itemize}}

\newenvironment{myenumerate}{%
\begin{enumerate}[leftmargin=1em, itemsep=.1em, parsep=.1em, topsep=.1em,
    partopsep=.1em]}
{\end{enumerate}}

\newcommand{\ie}{i.e.,\ }
\newcommand{\eg}{e.g.,\ }

\newenvironment{structure*}{\color{blue}\begin{myenumerate}}{\end{myenumerate}}

\sethlcolor{yellow}
\newcommand{\suggestion}[3][0em]{#3}
\newcommand{\update}[3][0em]{#3}

\newcommand{\myspecialcell}[2][c]{\begin{tabular}[#1]{@{}l@{}}#2\end{tabular}}

\begin{document}

\title[The 5W1H of Metadata Management]{Comprehensive and Comprehensible Data
Catalogs: \\The What, Who, Where, When, Why, and How of \\Metadata Management}

\author{Pranav Subramaniam$^1$, Yintong Ma$^1$, Chi Li$^1$, Ipsita Mohanty$^2$, Raul Castro Fernandez$^1$}
\affiliation{\vspace{0.1cm}
\institution{$^1$ University of Chicago, $^2$ University of Waterloo}
}
\email{{psubramaniam, yintongma, lichi, raulcf}@uchicago.edu, imohanty@uwaterloo.ca}

\input{sections/abstract.tex}

\maketitle

\input{sections/intro_v3.tex}
\input{sections/landscape.tex}

\input{sections/metadataDef.tex}

\input{sections/alt_schemacomparison_v2.tex}
\input{sections/evaluation_v2.tex}
\input{sections/relanddisc.tex}

\input{sections/conclusion.tex}

\newpage 
\balance 

\bibliographystyle{ACM-Reference-Format}
\bibliography{main}

\end{document}

%% file: sections/abstract.tex
\begin{abstract}
Data management tasks require access to metadata, which is increasingly tracked by databases called data catalogs.
Current catalogs are too dependent on users' understanding of data, leading to difficulties in large organizations of users with different skills: catalogs either make metadata easy for users to store and difficult to retrieve, or they make it easy to retrieve, but difficult to store.

 In this paper, we present 5W1H+R, a new catalog mental model
that is \emph{comprehensive} in the metadata it
represents, and \emph{comprehensible} in that it permits all users to locate metadata
easily. We demonstrate these properties via a user study.
We then discuss practical guidelines for implementing the new mental model.  We conclude mental models are important to make data catalogs
more useful and to boost metadata management efforts.
\end{abstract}

%% file: sections/intro_v3.tex
\section{Introduction}
\label{sec:introv3}

\begin{figure}[t]
    \centering
    \includegraphics[width=\columnwidth]{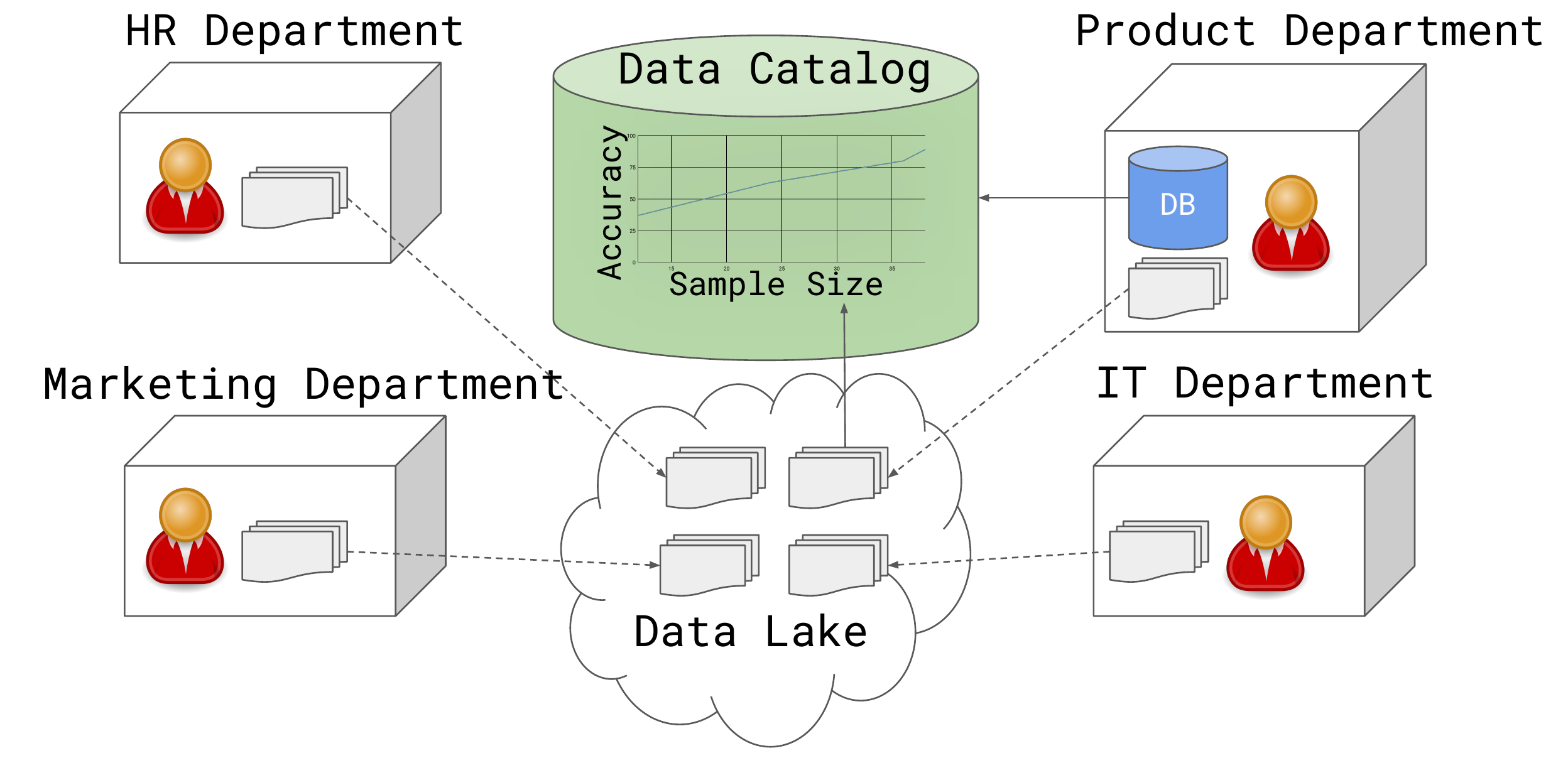}
    \vspace{-2ex}
  \caption{Organizations use data lakes to share data. Users may store data separately in their own databases. Data catalogs can share metadata, (e.g. model accuracy), for all lake \& DB data.}
\label{fig:mainprob}
\end{figure}

\begin{figure}[t]
  \centering
  \includegraphics[width=\columnwidth]{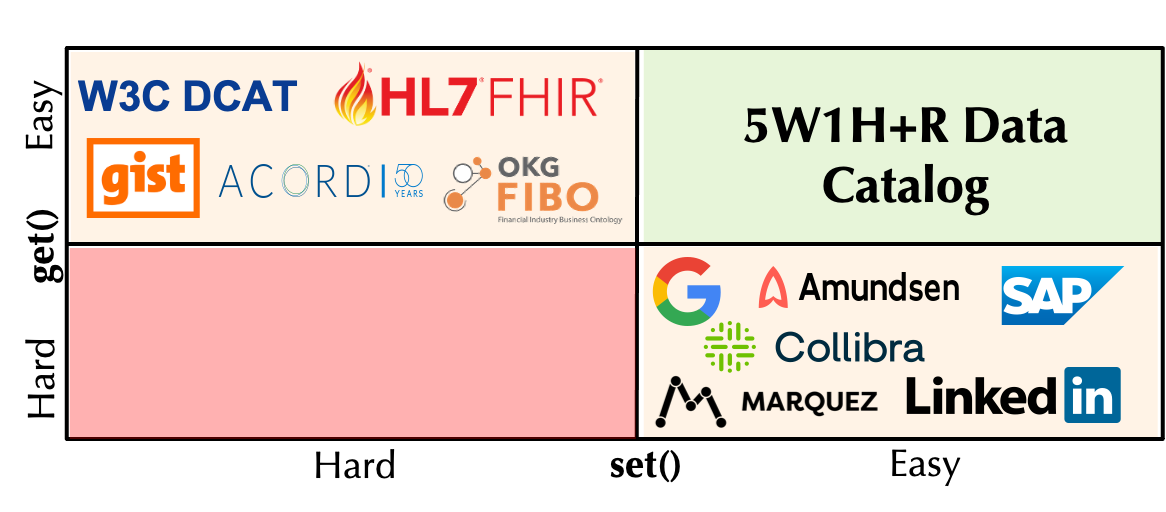}
  \vspace{-2ex}
  \caption{Design Space of Metadata Cataloging Solutions}
\label{fig:quadrant}
\end{figure}

In an increasingly data-driven world, a major data management challenge in
industry, the sciences, and the civic sphere, is the lack of solutions to find
relevant data (\textbf{D}iscovery), manage data usage and access
(\textbf{G}overnance), combine data to multiply its value
(\textbf{I}ntegration), and guarantee that data use is compliant with
regulations (\textbf{C}ompliance). A common denominator to all these
challenges is that addressing them requires access to \emph{metadata}:
information about the datasets that cannot be found in their contents. 

Many initiatives have appeared over the
last few years to organize and represent metadata to help diverse data users
address their data management tasks. The FAIR
principles~\cite{fairprinciples} in the sciences have galvanized academics to
complement data-sharing initiatives such as Dataverse~\cite{dataverse}, and
ICPSR~\cite{icpsr} with metadata that facilitates their usage. The explosion of
machine learning and its consequent pressure on data management has motivated
efforts to catalog data assets in the form of
datasheets~\cite{gebru2018datasheets} and enterprise metadata management
systems~\cite{gradflow, automatedvault}. We call software solutions designed to
store and retrieve metadata \textit{data catalogs}, which are databases for
metadata.

Many data management tasks greatly benefit from accessing a data catalog shared among a diverse group of data users who need to use each other's data. For example, \F\ref{fig:mainprob} illustrates a modern organization, where employees are organized in different units (e.g., departments, divisions) and do not necessarily know the data others own, even though they may need to use their data to solve data management tasks. These scenarios are challenging to navigate in the enterprise. Before using a dataset, employees must contact the data owners and understand the context before they can use such a dataset responsibly. It is precisely in these scenarios that a well-designed data catalog that centralizes metadata describing those datasets would be valuable. To be helpful, the catalog must be designed to provide a \emph{shared understanding of metadata} to all users so they know where to store the metadata and how to retrieve it, making it easy to use. If the catalog does not induce a shared understanding, then employees will ultimately ignore the catalog and be back to ground zero. Consider the following stylized example that illustrates the aforementioned problem:

\emph{Example.} Suppose Alice, a sales analyst, is in charge of predicting headphone sales for the incoming week, so the company can allocate the stock effectively. To build such a machine learning predictor, Alice searches for data she can use as a training dataset. Luckily, the company has recently set up a data catalog with a keyword search functionality. Using the search function Alice finds a possibly relevant dataset. Before committing to using this dataset, however, Alice, as a responsible engineer, wants to know a series of characteristics about the data that will ensure its use as part of the machine learning predictor is adequate and responsible. Alice would like to know: i) \emph{past use}: was this dataset used successfully for prediction in the past? If it was, this may give Alice additional context on who to ask for advice; ii) \emph{schema details}: whether the dataset contains relevant (and semantically meaningful) attribute names, and iii) \emph{creator details}: whether the dataset was created by a trustworthy data engineer. When Alice looks at the metadata listed in the catalog, she finds a plethora of metadata fields with possibly relevant names, such as \emph{logs}. However, Alice cannot tell whether these correspond to the information she is seeking. This semantic ambiguity naturally arises because of the distributed nature of the problem: those who populate the catalog may not use terminology that aligns with those who consume the information from the catalog.

Alice finds it difficult to successfully search using her desired metadata because other users are allowed to store metadata using a \textit{flexible schema}, which is provided by most enterprise data catalogs today. Different users may \emph{disagree} on the meaning of metadata, and the terms that should be used to describe it. Therefore, metadata may be named differently, or have different meanings, depending on the users who provide it. This makes it \emph{easy to store} metadata, but \emph{difficult to retrieve it}. Current enterprise catalogs can be characterized this way (see \F\ref{fig:quadrant}, bottom right). This can be alleviated if there is a \emph{shared mental model} (an agreed-upon set of terms) used for storage and retrieval of metadata. However, if this mental model is a \emph{large} ontology or controlled vocabulary such as W3C DCAT, or a manually constructed business dictionary, then retrieving metadata is easy, but storing metadata is difficult (\F\ref{fig:quadrant}, upper left), because (i) any user must correctly match their metadata to the agreed-upon term for the metadata the user has in mind. Also, (ii) the ontology may not be comprehensive: it must be modified if it does not include any user's metadata.

Therefore, a mental model must be: (1) \textbf{comprehensive} in the metadata it captures, (2) \textbf{comprehensible}: different users match the same metadata to the same term of the mental model (3) \textbf{easy-to-use}: the mental model should not have too many terms for users to match their metadata. A shared mental model with these properties would better facilitate data management tasks. Note, also, that such a mental model need not eliminate disagreements on metadata altogether (which may be undesirable, as different users may want to continue using their own terms). Rather, it should provide a shared set of terms to which all users can match their metadata in the same way, \emph{bounding disagreements} within each term.

In this paper, we propose a comprehensive, comprehensible, easy-to-use mental model based on the 5W1H principles of journalism~\cite{5w1href}.
\suggestion{R1O4}{ We choose 5W1H not only because of its desirable properties, but also because it is a natural choice, given the success of 5W1H-based description frameworks in several other computational contexts, ranging from describing user events that should trigger applications}~\cite{jang20055w1h}\suggestion{}{ to providing a complete description of next-generation probiotics}\cite{probiotics5w1h}.
It is informed by an in-depth study of the kinds of DGIC questions that appear in the literature. The new mental model facilitates the storage/retrieval of metadata to technical (data scientists/engineers) and non-technical (business analysts, sales teams) users alike. We illustrate how this mental model can be implemented in a catalog by outlining practical guidelines for a catalog that preserves the mental model's \emph{cognitive fit} \cite{cognitivefitDef} with user's metadata specifications.

We conduct an IRB-approved user study to demonstrate the mental model benefits and compare the new mental model with those implicitly defined in existing catalogs, namely, Google Cloud Service's (GCS) data catalog \KR, and LinkedIn's Datahub \KR, which we can derive from the schemas they expose to users. Our results show that the new mental model makes it easier for users to store and retrieve metadata consistently, meaning that it can reduce the effort needed to find the metadata required for DGIC tasks. Our results further demonstrate this reduction of effort through difficulty ratings, that show that our mental model is easier to use. Further, our mental model has fewer cases where users state they do not know how they would use the mental model to find metadata, providing further evidence that our mental model is comprehensive.


\mypar{Scope of this paper} The contribution of this paper are:

\begin{myitemize}

\item A study of the data catalog landscape and metadata needs (Section
\ref{sec:landscape}).

\item A comprehensive and comprehensible \textbf{mental model} (Sections \ref{subsubsec:individual} and
\ref{subsubsec:many}).

\item A set of practical guidelines for implementing the mental model in a
catalog in a way that preserves comprehensiveness and comprehensibility (Section
\ref{subsec:altcatalogv2}).

\item A user study that demonstrates the benefits of well-designed
mental models for metadata management (Section \ref{sec:evaluationv2}).

\end{myitemize}

\mypar{Out of scope topics} This paper also has some non-goals:
\begin{myitemize}

\item This paper does not propose a full-fledged data catalog.
We claim that
\emph{any} catalog benefits from well-designed mental models and we demonstrate
that by comparing other catalogs' mental models to the new one we propose.
Consequently, we do not address all the practical concerns associated with
implementing a data catalog.  

\item A UI can help catalog users store and retrieve metadata. We do not discuss
the problem of designing UIs that expose a catalog mental model in this paper,
and we do not discuss the UX/IX challenges associated with this.

\end{myitemize}

\noindent\textbf{Metadata for Data Management.} \suggestion{R2}{Metadata management solutions store metadata before a user performs any data management task. Users can use metadata to better solve their data management tasks, either by enabling construction of context-specific data management solutions, or by better using the outputs of existing data management solutions.  An example of the former: if a table description indicates that all instances of geospatial data were collected in the United States, then an analyst's preprocessing step may be to remove data instances whose latitudes/longitudes do not correspond to United States locations. An example of the latter: using stored table descriptions for a data repository, a user can determine which overlap similarity-based joins discovered in the repository are sensible, according to analysts' domain knowledge.}

Users face substantial overheads in solving their data management tasks due to their varying domains and types of expertise, which make it difficult for them to understand metadata provided by others. The lack of metadata management solutions that handle this human aspect, and not just the system scalability aspect, hampers data management efforts. 
We have
informed the motivation and contributions of this paper by consulting with a
large number of industry and science stakeholders, and we make an effort in
connecting our work to other initiatives with the aim of contributing to
metadata management academically and practically.
We show that this representation is more comprehensible and easy to use compared to those of enterprise catalogs.

\mypar{Outline} Section~\ref{sec:landscape} explains the
characterization of data management questions into the DGIC categories, as well
as the ecosystem of data catalog and cataloging solutions. We introduce our
mental model and guidelines for its implementation in Section~\ref{sec:metadata}. Section~\ref{sec:evaluationv2}
presents evaluation results, followed by related efforts and related work in Section~\ref{sec:relanddisc}, and conclusions in Section~\ref{sec:conclusions}.

%% file: sections/landscape.tex
\section{Landscape of Data Catalogs}
\label{sec:landscape}

Consider Alice, the sales analyst from the introduction. In order to find training data for sales prediction, she needs \emph{past use}, \emph{creation details}, and \emph{schema details} metadata.

In the example, sentences in italics indicate intermediary tasks that need to be achieved to address the goal.
Each intermediate task can be
assisted with relevant metadata, such as a dataset annotation explaining its \emph{past} use--this would make it easier for Alice to determine if the dataset should be used for training machine learning models. Without this metadata available, solving each
of these tasks is usually a time-consuming, tedious, and ill-defined process
that consumes much of data workers' energy because they may need to identify and chase the right person in the organization with the answers to their questions (that is, their required metadata). 

Making that metadata available
requires: i) creating the metadata and, ii) representing metadata in a way that
is useful to other data workers. Metadata can be created automatically, via
software, or manually, and a catalog must accommodate both. In this paper we are
primarily concerned with ii): how to represent metadata.

\input{tables/dgicquestions.tex}

We define metadata and mental models in Section~\ref{subsec:metakr}, we discuss
the metadata landscape in Section~\ref{subsec:metaland}, and conclude with a
discussion of current data catalogs in Section~\ref{subsec:dgicsupport}.

\subsection{Metadata and Mental Models} 
\label{subsec:metakr}

\textbf{Data assets} are artifacts describable with metadata, including relational tables, such as tables in databases, spreadsheets, or CSV files, rows or columns of these tables, unstructured files, derived data products, others, or groups of any of these. 
A \textbf{metadata-item (MI)} is a \textsc{key:value}
pair where the \textsc{key} indicates a property of interest and \textsc{value}
contains its value. The property of interest describes a
\emph{data asset} or a group of \emph{data assets}.
MIs are stored in a \textbf{catalog} via \textsc{set(key, value)} operations and
retrieved using a \textsc{get(key)} operation. A catalog stores the MIs that refer to data assets.



Applications may describe the same property of a data asset differently, leading
to different MIs. Or they may express the MI values using different
representations, \eg different units. This flexibility lets applications
describe data assets without constraints but makes retrieving metadata
challenging because the MI key is not known a priori. 

A \textbf{mental model} (\KR) is a partitioning of the metadata items $C$ into
subsets $\{p_1, \ldots, p_n \in P | p_i \subseteq C, p_i \cap p_j = \emptyset,
\forall i, j\}$, where $\bigcup_{i=1}^n p_i = C$.  Without a \KR\ available,
retrieving a MI without knowing its key requires searching through all MIs. If a
\KR\ is available, retrieving a MI requires i) finding the $p_i$ where it
belongs, and ii) searching through all MIs in $p_i$. Mappings between
MIs and \KR's partitions do not exist a priori and depend on users' and
applications' understanding of both MIs and the \KR's partitions, \ie users of the
\KR\ decide where to map a MI. A good \KR\ is one that leads two users to agree,
on the partition where a MI belongs without explicit communication. Hence,
agreeing on a mental model leads to consistent decisions even when MM users in different teams have different technical skills, use different terminology, and/or do not communicate with each other.

\mypar{Quality Metrics of a \KR} A \KR\ is \textit{comprehensive} in the
metadata it describes if every MI can be placed in a partition of the \KR. A
\KR\ is \textit{comprehensible} if it generates agreement on the mapping between
MIs and partitions among users of a catalog implementing that \KR.  A \KR\ is
\textit{easy to use} if \KR\ users find it easy to map MIs to partitions.  Many
catalogs include a \emph{catch-all partition} meant to capture metadata that
does not fit in any other partition. Although \emph{catch-all partitions}
increase the comprehensiveness of a mental model, they reduce its
comprehensibility because any metadata applies to this partition. We account for
\emph{catch-all} partitions in the evaluation section by considering a mapping
to this partition as a \emph{None} response, that indicates no mapping exists.


%

\subsection{The Metadata Landscape} 
\label{subsec:metaland}

In practice there are many possible MIs used to facilitate data management tasks. To gain intuition on those commonly
used we have surveyed \suggestion{R3O1}{research about various data-intensive applications such as DBMSs and machine learning literature, data curation research and efforts, and industry data catalog use cases. We have also} consulted with data employees. We have
extracted 155 DGIC questions data workers often face that would be addressed
with access to the right MIs.  These questions illustrate the metadata
landscape we consider in this paper. By grouping redundant or highly similar questions (\eg questions that were more specific versions of other questions) together, we have summarized the 155 questions into
27, shown in Table~\ref{table:commondataquestions}. \update{R3D4}{Note that classifying a question means identifying a partition where its answer MI should be stored.}

\mypar{The DGIC questions} We do not claim our DGIC definitions are universal,
but we find them useful for explaining the kinds of data questions we find in
the literature, and our contribution (our \KR) does not depend on them. We
synthesized the definitions from multiple sources: \textit{\textbf{D}iscovery}
involves selecting a subset of datasets from a larger set to satisfy specific
criteria \cite{aurum, goods, collibra_main}. \textit{\textbf{G}overnance}
involves ensuring that the purpose of a data asset, and the method used to
fulfill that purpose are understood by all data users. Note that our definition
of governance is more inclusive than existing definitions \cite{datagovdefs},
which consider governance tasks to be related to decision rights and
accountabilities regarding who uses data, and for what.
\textit{\textbf{I}ntegration} involves combining existing data assets, which may
require preparation \cite{biggorilla, civilizer}.  \textit{\textbf{C}ompliance}
involves ensuring that data assets with sensitive information meet regulatory
requirements \cite{gdprcomp, ccpacomp}. We show in
Table~\ref{table:commondataquestions} the question classification into DGIC.

\subsection{Analysis of Today's Data Catalogs}
\label{subsec:dgicsupport} 

We study existing data catalogs to understand what features they provide to
store and retrieve metadata. The results are shown in
Table~\ref{table:dgictable}. \update{R2D3}{}\textit{Keyword} \update{}{(W) means the catalog provides a
full-text search feature to search for metadata. } \textit{Key} \update{}{(K) indicates the
catalog explicitly provides a pre-specified collection of MI keys, e.g., 
when
the catalog provides a collection of MI keys which one can search through.}
 \textit{Partition} (P) indicates the
catalog provides a partition of a \KR\ to classify a MI (\eg by providing a field where all metadata of a type can be added).

\input{tables/catalogcomparison.tex}

Some catalogs were designed to address different areas of DGIC so
have more support for those than others. Although we are not aware if catalogs
were designed around a \KR\ explicitly, we can derive their \KR\ from their
documentation by observing what users need to know to store and retrieve different MIs. 
All the analyzed catalogs implement a \KR\ consisting of: (i)
partitions for an explicitly specified set of \textsc{key:value} pairs of
metadata; and (ii) a \textit{catch-all partition}: a single partition for all
\textsc{key:value} pairs the catalog does not describe. Catalogs provide this catch-all partition not only to allow for the tracking of metadata it does not capture explicitly, but also to allow users to define their own domain-specific (or tribally-known) terms (\eg for use among team members).

While the catch-all partition allows these \KR s to be comprehensive in the MI
they describe, they lead to lack of comprehensibility because they accept any
MI (Section \ref{subsec:metakr}). Therefore, these catalogs sit in the bottom right
corner of \F\ref{fig:quadrant}: it is easy to store MIs but difficult to find them later, especially for other users who did not store these MIs.

On the other hand, controlled vocabularies such as W3C DCAT consist of many
partitions, each with a single \textsc{key:value} pair. This makes it easy to
query data catalogs implementing this ontology as a \KR\ as long as the terms in
the vocabulary are known by the user who submits the query. However, it is
difficult to represent metadata with these catalogs because each MI needs to be
annotated with one of the many partitions. As a consequence, we classify
controlled vocabularies in the top left corner of the \F\ref{fig:quadrant}
quadrant.
We discuss how to leverage ontologies as part of data catalogs in Section \ref{sec:discussion}.

%% file: tables/dgicquestions.tex
\begin{table*}[]
\centering
\small
\begin{tabular}{@{}llll@{}}
\toprule
 &
  Representatives of Common Data Questions &
  DGIC Category &
  5W1H+R Partition \\ \midrule
Q1 &
  For what purpose was the dataset created? &
  G \cite{gebru2018datasheets, alation_guidetocatalog} &
  Why \\
Q2 &
  \myspecialcell{Are there tasks for which the dataset should not be used?} &
  G,C \cite{gebru2018datasheets} &
  Why \\
Q3 &
  Who created the dataset? &
  G,C \cite{gebru2018datasheets, halevy2016goods} &
  Who \\
Q4 &
  Who was involved in the data creation process? &
  G,C \cite{gebru2018datasheets} &
  Who \\
Q5 &
  \myspecialcell{How can the owner/curator/manager of the dataset be contacted?} &
  G \cite{gebru2018datasheets} &
  Who \\
Q6 &
  \myspecialcell{What are the privacy and legal constraints on the accessibility of the dataset?} &
  C \cite{lawrence2017data} &
  Who \\
Q7 &
  Is there an access control list for the dataset? &
  G,D \cite{halevy2016goods} &
  Who \\
Q8 &
  What is the reputation of the creator of a dataset? &
  G \cite{gregory2019lost} &
  Who \\
Q9 &
  What do the instances of the dataset represent? &
  D,G,I \cite{gebru2018datasheets} &
  What \\
Q10 &
  What is the size of the dataset? &
  D,G,I \cite{halevy2016goods} &
  What \\
Q11 &
  Are there errors in the dataset? &
  D,G,I \cite{gebru2018datasheets, gregory2019lost, lawrence2017data} &
  What \\
Q12 &
  Does the dataset have missing values? &
  D,G,I \cite{gregory2019lost} &
  What \\
Q13 &
  What is the domain of the values in this dataset? &
  D,G,I \cite{holland2018dataset} &
  What \\
Q14 &
  \myspecialcell{If the dataset is a sample of a larger dataset, what was the sampling strategy?} &
  G,I \cite{gebru2018datasheets} &
  How \\
Q15 &
  \myspecialcell{Does the dataset contain personally identifiable information (PII)?} &
  G,C \cite{alation_guidetocatalog, bapat_2020} &
  What \\
Q16 &
  What is the quality of the dataset? &
  G \cite{bapat_2020, collibra_main, lawrenz2020significant, atlas_docs} &
  What \\
Q17 &
  \myspecialcell{Was any preprocessing/cleaning/labeling of the dataset done?} &
  G \cite{gebru2018datasheets} &
  How \\
Q18 &
  \myspecialcell{Was data collection randomized? Could it be biased in any way?} &
  G \cite{lawrence2017data} &
  How \\
Q19 &
  \myspecialcell{Is there anything about dataset preprocessing/cleaning that could impact future uses?} &
  G \cite{gebru2018datasheets} &
  How \\
Q20 &
  What is the dataset’s release date? &
  D,G,I \cite{holland2018dataset} &
  When \\
Q21 &
  Is there an expiration date for this dataset? &
  D,G \cite{atlas_docs} &
  When \\
Q22 &
  How often will the dataset be updated? &
  G,I \cite{gebru2018datasheets} &
  When \\
Q23 &
  When was the data last modified? &
  D,G,I \cite{halevy2016goods} &
  When \\
Q24 &
  How easy is it to download and explore this dataset? &
  D \cite{gregory2019lost} &
  Where \\
Q25 &
  \myspecialcell{What is the format of the dataset, and what type of repository is the dataset located in?} &
  D \cite{lawrence2017data} &
  Where \\
Q26 &
  What is the provenance of this dataset? &
  I \cite{zhang2020finding} &
  Relationship \\
Q27 &
  \myspecialcell{What other datasets exist in this repository that are related to this dataset?} &
  D,G,I \cite{yakout2012infogather} &
  Relationship \\ \bottomrule
\end{tabular}%
\caption{Questions Users Ask about Data, their DGIC Classifications, and 5W1H+R Partitions}
\label{table:commondataquestions}
\end{table*}

%% file: tables/catalogcomparison.tex

\begin{table}
\centering
\begin{tabular}{@{}lcccc@{}}
\toprule
                               & D & G & I  & C \\ \midrule
Berkeley's Ground~\cite{hellerstein2017ground}              & K & P & P  & P \\
Microsoft Azure Data Catalog~\cite{azure_common}   & W & P & P  & P \\
Apache Atlas~\cite{atlas_docs}                   & W & P & P  & P \\
Denodo platform~\cite{denodo}                & W & P & P & P \\
SAP Data Intelligence platform~\cite{sap_main} & W & P & K  & K \\
Boomi Data platform~\cite{boomi}            & W & P & P & P \\
WeWork's Marquez~\cite{marquez}               & K & P & P  & P \\
Lyft's Amundsen~\cite{amundsen}                & W & P & P  & P \\
Ataccama Data Catalog~\cite{ataccama}          & K & K & P & K \\
Atlan Data Catalog~\cite{atlan}                & W & P & P & P \\
Linkedin's Datahub~\cite{datahub}             & W & P & P  & P \\
Informatica's Data Catalog~\cite{informaticadatacat} & W & P & P & K \\ \bottomrule
\end{tabular}%
\captionof{table}{DGIC support in existing catalogs. \textit{Keyword} (W),
\textit{Key} (K), \textit{Partition} (P).}
\label{table:dgictable} 
\end{table}

%% file: sections/metadataDef.tex
\section{The 5W1H+R Mental Model}
\label{sec:metadata}

In this section, we introduce the new 5W1H+R mental model. The 5W1H component of
the \KR\ applies to MIs of individual data assets
(Section~\ref{subsubsec:individual}) and the +R component applies to
relationships between more than one data asset (Section~\ref{subsubsec:many}).
We explain why the \KR\ is comprehensive, comprehensible and easy to use in
Section~\ref{subsubsec:whythiskr}. We conclude by offering an implementation
guide of the \KR\ in a catalog (Section~\ref{subsec:altcatalogv2}). This guide
illustrates that it is possible to implement the \KR\ in a catalog without losing its
properties.

\subsection{5W1H-Profiles}
\label{subsubsec:individual}

Given a MI that describes one data asset, the MI fits into one of the 5W1H
partitions of the \KR. We have included a column with the 5W1H partition in
Table~\ref{table:commondataquestions} to explain where the MI useful to answer
those questions fits. Note that we call each 5W1H partition of a data asset a
\textit{profile}.

A mnemonic for the 5W1H \KR\ is the following: "\textbf{Why}, \textbf{When}, and
\textbf{How} does \textbf{Who} use a \textbf{Data Asset} (located
\textbf{Where}) whose contents are described by \textbf{What}?"

\mypar{Who-profile} Items identifying persons or software that created,
modified, or can access the data asset, and/or explaining their relationship
with the data asset (e.g. role information, access privileges to the data
asset). A data question falls in this partition if: i) the question can be
answered by information about who used the data asset before; ii) the
question can be answered by information about who can access the asset.
For example, the contact information of a data owner would belong in this partition. 

\mypar{What-profile} Information that requires looking at the data. A 
question falls in this partition if: i) it can be
answered by reading the data directly and/or performing computations on the
data. ii) it can be answered through (existing) semantic information
about the data, such as schema annotations or descriptions.
For example, the most frequent product name that appears in a relational table of customer transaction data would belong in this partition. 

\mypar{When-profile} Temporal information about the data asset lifecycle: when it was
created, modified, when does it expire, etc. A data question falls in this
partition if: i) the question can be answered by information
about data asset usage during a particular time period; ii) the question can be
answered by information about when the data asset is available.
For example, the frequency with which a data asset is updated would belong in this partition.  

\mypar{Why-profile} Explains why the data asset exists, its purpose, and intended
use. A data question falls in this partition if: i) the question can be answered
by information about why the data asset was used a certain way, including why the
data asset was created and deleted; ii) the question can be answered by information
related to the intended uses of the data asset.
For example, the specific research question a data asset was designed to answer \cite{maodsdomainexperts} would belong in this partition.

\mypar{Where-profile} Physical location of the data asset and information about how
it is accessed. A data question falls in this partition if: i) the question can
be answered by information about how to access this data asset; ii) the question
can be answered by information about the format of the data asset being stored
(such as CSV file vs. postgres table vs. mysql table).  iii) the question can be
answered by information about the source where this data asset can be located.
For example, the name of the database where the data asset is stored would belong in this partition.  
    
\mypar{How-profile} Information about the processes that produced, modified, or
read the data asset. This includes data collection and preparation methods, as well
as programs, queries, or artifacts that were run on the data asset. A data question
falls in this partition if this question be answered by information about how a
data asset has been used (read, modified, or created) for a task/application.
For example, the script used to remove missing values from a CSV file so that accurate statistical analysis can be performed would belong in this partition.  

\subsection{Relationships Profile}
\label{subsubsec:many}

When a MI refers to more than one asset we do not use a profile, but a
relationship. Examples of such MIs that are already considered useful for facilitating
discovery and integration come from the current data management literature on
metadata extraction: the Jaccard similarity between 2 columns as data assets and
containment (inclusion dependencies) between columns are used to guess whether
the entities two columns represent are the same. This is used in data discovery
systems such as Aurum \cite{fernandez2018aurum} to allow users to search for
columns that are similar.  Another example is a link or reference to a dataflow
graph describing \textit{provenance} (the set of operations used to derive one
data asset from another).  This can be used to guess how raw data has been
processed, and what processing may still be needed. This is also used in data
discovery systems, such as Juneau \cite{zhang2020finding}, to allow users to
search for tables that are likely to have been generated by performing
operations on a table the user is familiar with.
Other examples come from
literature on collaboration between data scientists, such as the team
responsible for creating/maintaining multiple data assets
\cite{koestenstructured}, or the specific tool/script used to run analysis on a
group of data assets \cite{maodsdomainexperts}. 

The \KR\ includes a \emph{relationship} partition (+R) to describe these MIs.  As
with individual items, it is useful to think about the types of relationships
the \KR\ should support in terms of the 5W1H partitions.  For example, consider
2 columns as data assets. Their Jaccard similarity and containment (inclusion
dependency) is a relationship on their What-Profiles because both can be
computed by looking at the data.  In fact, any properties that can be computed
by reading one or more assets, such as Jaccard similarity and containment, are
relationships on the What-Profiles of all assets involved.  The 5W1H partitions
can describe not only these types of metadata used to describe
relational properties and similarities between columns and tables, but also
other types of relationships. A PK/FK relationship can be
represented as a Why-Profile if the table PK/FK columns were produced with the
purpose of indicating a join relationship between two relations. 
Consider
relations A and B, where B is a version of A. We can express their temporal
relationship using a When-Profile, their provenance using a How-Profile, etc. We
can relate assets according to who created them, or who accessed them using
relationships on Who-Profiles. And we can store relationship MI as combinations
of profiles of related assets.

\emph{Example.} Consider our example from Section~\ref{sec:introv3}. If Alice could not only examine \emph{past use} of one dataset, but find all datasets that are \emph{related in that they were used to train the same ML model} (related on how-profiles), \emph{related in that they were created by the same trustworthy data engineer} (related on who-profiles), and \emph{related in that all datasets contain relevant attribute names} (related on what-profiles), Alice's search is much easier than if she tries to view each individual dataset to see if it was used for training.

\subsection{The 5W1H+R \KR\ Rationale}
\label{subsubsec:whythiskr}

The 5W1H+R \KR\ \emph{bounds disagreements} when setting/getting MIs by
presenting users with a \KR\ with high cognitive fit, \ie one that leads users
to implicitly agree on the mapping of a MI to a \KR's partition. 
On choosing the
\KR\ we strived to make it comprehensive, comprehensible, and
easy-to-use:


\begin{myitemize}
\item \textbf{Comprehensive.} Journalists use 5W1H to cover news because it
leads to a broad coverage of the event. We could represent all 155 DGIC
questions extracted from the literature in the \KR. We include the 5W1H+R
partition along with the 27 questions that synthesize the corpus in
Table~\ref{table:commondataquestions}. 

\item \textbf{Comprehensible.} The 5W1H+R \KR\ has high cognitive
fit~\cite{cognitivefitDef} with any task involving the storage or retrieval of
metadata. A \KR\ has cognitive fit with a task if there is a common mapping
among users performing the same task between a \KR's \emph{internal
representation} (the way the \KR\ partitions metadata) and the \emph{external
task representation} (a user's specification of the metadata required).

\item \textbf{Easy-to-use.} The 5W1H+R \KR\ is explained in
natural language. Most natural languages describe objects and events using 5W1H,
making it a familiar set of terms to most people, technical and non-technical
alike.  

\end{myitemize}
\noindent \suggestion{R1O3}{A MM with the above qualitative properties partitions metadata such that the overall communication effort of storing or retrieving metadata is low.}
We provide quantitative data substantiating the above, obtained
through a user study, in Section~\ref{sec:evaluationv2}.

%% file: sections/alt_schemacomparison_v2.tex
\subsection{Guide to Mental Model Implementation}
\label{subsec:altcatalogv2}

The 5W1H+R \KR\ is comprehensive and comprehensible, but one may implement it in
a catalog in a way that loses these properties.  In this section, we briefly
illustrate how the 5W1H+R mental model can be implemented in a catalog without
losing its comprehensiveness and comprehensibility by explaining the following
practical guidelines for such a catalog, and explaining how current data
management technologies can be used to implement them.

\mypar{G1. Maintain cognitive fit} The catalog conceptual schema should
explicitly represent the 5W1H+R profiles and relationships to maintain high
cognitive fit (high comprehensibility). For example, in the catalog database
schema, there should be a table for each 5W1H profile and relationships.  If
this guideline is not provided by the catalog, then users will map their metadata
to other entities either before or instead of using the 5W1H+R partitions.  If a
user maps their metadata to another entity before using the 5W1H+R \KR, then
they must be able to map the entity into one 5W1H+R partition, which may not be
possible (\eg the entity consists of a user, the script run by the user, and the
file repository where a data asset belongs).  If a user maps their metadata to
another entity instead of using 5W1H+R, then they lose the comprehensibility
property that the 5W1H+R \KR\ provides. In both these cases, the user
potentially loses the comprehensibility of the 5W1H+R \KR. 
Common schema design methodologies used in data management such as
snowflake schemas\cite{starschema_adv} and data vault\cite{datavault_def} accommodate this guideline.

\mypar{G2. Preserve history} At any point, different users may be producing MIs
about the same data asset. The catalog must represent all versions of MIs
referring to the same data asset, and let the consumer of that metadata
reconcile or decide which one to trust. This means that operations on the
catalog are append-only and that updates should only be allowed in special
circumstances, \ie when a user wants to correct a mistake. For example, a user
may choose the word "uniques" to describe the number of unique values in a
relation column but express the quantity with a ratio, while a software profiler
may choose "unique\_number\_values " and store the absolute number instead.
It is not the responsibility of a catalog to decide whether one of these names
is preferred--the catalog implements a \KR\ to bound disagreements (as opposed
to eliminating them). The catalog should expose both versions to consumers, who
may decide to run software that indicates statistical similarity above a
threshold, resolve the ambiguity manually, or live with different definitions.
To decide, consumers not only have the different versions available, but also
the audit metadata associated with each entry. This brings us to our next
guideline. 

\mypar{G3. Track audit metadata} Audit metadata is information about the
\textsc{set} and \textsc{get} operations users perform on metadata in the
catalog.  Such audit metadata is stored in order to help catalog users better
track and understand their metadata so they can perform tasks such as tracking
how a property of a data asset evolved over time, or, as explained by (G2), they
can reconcile differences in MIs that share the same key, but have different
values (\eg by trusting the MI with the later timestamp).  A catalog schema must
represent, in addition to the mental model, audit metadata that includes who
(what user) is setting a MI, the version of the MI (to differentiate among
existing ones), and the time at which the MI was stored. Many catalogs track
this audit metadata already \cite{datahub, amundsen}, and even if they do not,
there are event streaming solutions \cite{kafka} that can facilitate the
tracking of audit metadata as well. 


%
%

\mypar{G4. Support for complex metadata} MIs can be represented in many
different formats: primitive types may be sufficient to describe some MIs (\eg
the average of a numerical column of a CSV file), but other MIs may be better
represented with more complex data types (\eg the list of users who maintain a
data asset).  The catalog should accommodate complex values for metadata items
so it must not impose a fixed primitive type. We can represent MIs as
\textsc{key:value} pairs, which can be efficiently supported in modern database
systems with a JSON type. This permits storing arbitrary objects in the
\textsc{value}, including URLs to externally stored metadata.


A catalog implementing the above 4 guidelines, which can be obtained through existing
technologies, can implement the 5W1H+R \KR\ or any other comprehensive
comprehensible \KR. Specifically, such a catalog represents the \KR\ profiles
and relationships in its schema (\eg by creating tables for each profile)
(\textbf{G1}), is capable of uniquely identifying multiple versions of the same
MI (\textbf{G2}) using audit metadata (\textbf{G3}) (\eg by using event
streaming to track audit metadata), and the MIs can be represented with a type
compatible general \textsc{key:value} representation (\eg as JSON objects)
(\textbf{G4}).

%% file: sections/evaluation_v2.tex
\section{Evaluation}
\label{sec:evaluationv2}

In this section, we present the evaluation results. A good metadata mental model
is comprehensive, comprehensible, and easy to use. The
5W1H+R MM is comprehensive: we could represent metadata items (MIs) answering
all 27 questions (see Table~\ref{table:commondataquestions}) that were
synthesized from 155 questions found in diverse literature, blog posts, and
reports from data users. Then, in this
section, we evaluate whether the mental model is comprehensible and easy to use.
To achieve that we conduct a user study where we measure the comprehensibility
and ease-of-use of the 5W1H+R MM, and compare it with that of Google's\cite{googletechvsbus} and
LinkedIn's\cite{datahub} catalog solutions. We first introduce the mental models we use in the
user study in Section~\ref{subsec:targets}. We follow with the design of the
user study in Section~\ref{subsubsec:usdesign} and present the results in
Section~\ref{subsec:consistency}.



\subsection{Target Mental Models}
\label{subsec:targets}

We compare the 5W1H+R \KR\ with those of LinkedIn's Datahub\cite{datahub} and Google Data
Catalog (GCS)\cite{googletechvsbus}. Datahub and GCS are good representatives of both open-source
catalogs such as Amundsen\cite{amundsen}, Marquez\cite{marquez}, Apache Atlas\cite{atlas_docs}, and cloud service catalogs
such as Microsoft Azure Data Catalog\cite{azure_common}, Denodo Cloud Platform\cite{denodo}, and SAP Data
Intelligence\cite{sap_main}. Due to the increasing interest in data catalogs, new offerings
appear frequently~\footnote{an industrial collaborator tells us they have tested
~40 different catalogs}, but these catalogs remain a good representation of
today's offerings.

Today's catalogs do not have an explicitly designed mental model. Instead, data
users learn the schemas exposed by these catalogs, \eg via the UI or other APIs,
and use those to store and retrieve metadata. Therefore, to extract a
catalog mental model, we designed a rubric. Then, 3 of the authors applied
the rubric to each of GCS and Datahub independently. Finally, we derived those
catalogs' mental model from the output of applying the rubric, and we used them
for the study. 

\mypar{Rubric} The goal of the rubric is to provide guidelines to determine the list
of metadata entities and attributes that a catalog exposes to data users. The
rubric is as follows:


\begin{figure}[!htb]
\vspace{-1mm}
\centering
    \includegraphics[width=\columnwidth]{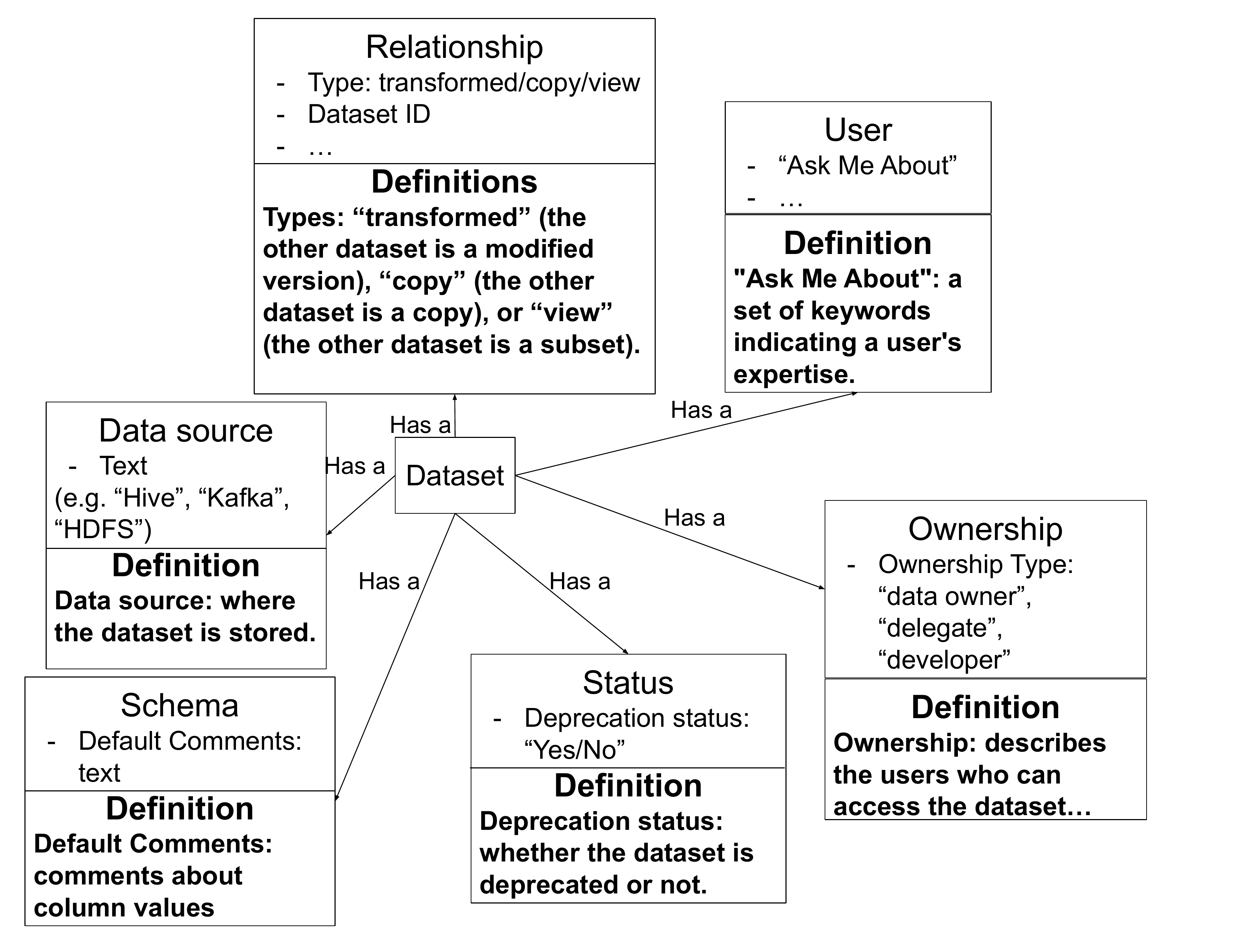}
    \caption{Datahub MM Partition Names and Definitions}
    \label{fig:datahubMM}
\end{figure}

\mypar{Step 1} Each author deploys the
target catalog and inserts a data asset (\ie table or spreadsheet) into the
catalog. In the process of doing so, the author: (i) records the terminology
provided by the catalog's UI when inserting the asset (\eg the names of the text
fields for entering metadata). Each recorded term $X$ is an entity: a category
of information about a data asset.  We say a data asset "has $X$". (ii) Record terminology the data catalog generates for a
data asset upon insertion (\eg automatically-generated five-number summaries of
numerical columns).  (iii) Record any constructs for inserting further metadata a
user may want to provide for a data asset. (\eg a "tags" page). Such constructs
are catch-all partitions where users can store/retrieve any metadata the
catalog's schema does not explicitly represent.

\mypar{Step 2} Each author examines how the catalog represents the inserted
asset. They record the titles of groups of entities provided about an asset
(\eg "Properties", or "Documentation"), as well as entities themselves.
Groups normally come in the form of tabs, or headings.  We say an asset
"has" $G$, where $G$ is the title of a group of entities, and $G$ in turn "has"
entities $X$ as its attributes.

\begin{figure}[!htb]
\vspace{-1mm}
\centering
    \includegraphics[width=\columnwidth]{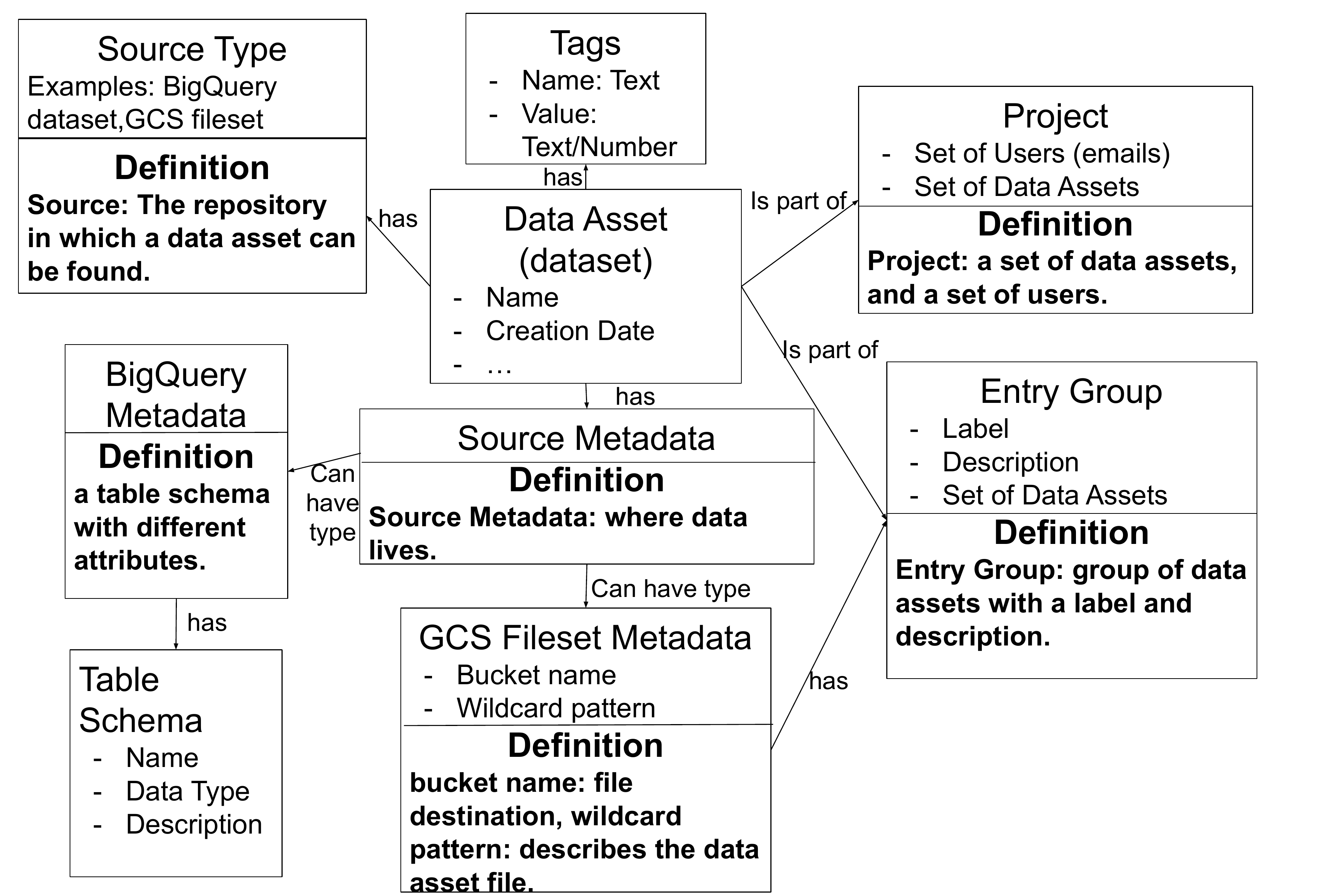}
    \caption{GCS MM Partition Names and Definitions}
    \label{fig:gcsMM}
\end{figure}

\mypar{Step 3} Each author represents the MM using the list of entity-attribute
statements that are of the form "$N$ has $X$". They start with statements of the
form "A data asset has $G$". These groups $G$ represent the highest-level
distinction a catalog makes about metadata. Therefore, the titles of these
groups $G$ are the \emph{partitions} that form the catalog's MM.

\begin{figure}[!htb]
\vspace{-1mm}
\centering
    \includegraphics[width=\columnwidth]{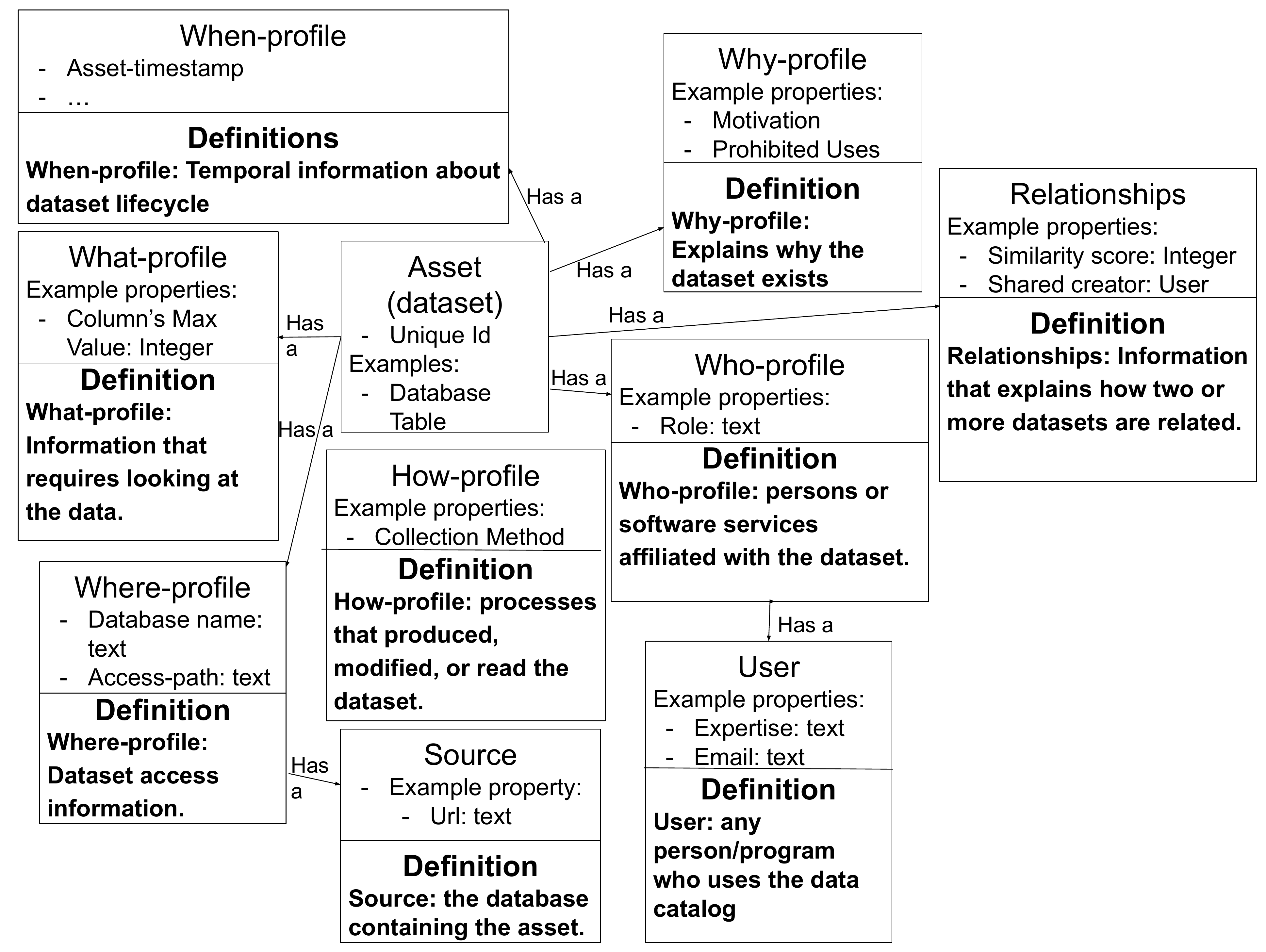}
    \caption{5W1H+R MM Partition Names and Definitions}
    \label{fig:5W1HMM}
\end{figure}


An existing catalog MM consists of the highest level distinctions about
metadata exposed to users. Figures \ref{fig:5W1HMM},
\ref{fig:gcsMM}, and \ref{fig:datahubMM} display simplified versions of the \KR s of 5W1H+R, GCS, and
Datahub respectively, shown to users. We have simplified them for readability.

\subsection{Between-Subjects Study Design, Participant Recruitment, and Threats
to Validity} 
\label{subsubsec:usdesign}

We implement the user study as a survey distributed via
Prolific~\cite{prolific} to 160 participants. The high-level flow of the study
is as follows. First, we present participants with the questions from
Table~\ref{table:commondataquestions} and ask them to rank familiarity
with those questions. We only ask them questions about those questions they are
very familiar with: this is how we ensure they are data users. Second, we
present them one of the three \KR, and make sure they were not familiar with the
\KR\ a priori to avoid confounding effects. Then, for each set of questions they
claimed familiarity with and the assigned \KR, we gather results on the
two metrics of interest as follows. 



\noindent\textbf{Consistency: a measure of comprehensibility} Each participant
classifies the question to a \KR 's partition. If they find no partition is
adequate, they choose \textsc{None}, (analogous to selecting the
catch-all partition in GCS or Datahub). 
\suggestion{R3O2}{As an example of this classification, consider the MI indicating whether a dataset contains PII. A user may classify this as a \textit{What-profile} according to the 5W1H+R MM, since one needs to look at the data to determine this. A user may classify this as a \textit{Table Schema} according to the GCS MM and \textit{Schema} according to the Datahub MM, as schema details need to be known in order to determine whether any table attribute contains PII. }
Participants' partition choices for a
question provide a frequency distribution over the partitions of a \KR. We use
this frequency distribution to measure the extent to which participants chose
the same partition (or the same few partitions) for a question. Concretely, if
most participants classified a question into one partition, that indicates there
is an agreement among users on where a MI belongs. Contrast that with an
inconsistent \KR\, that would cause users to be divided on which partition to
choose in the worst case; the worst case being an even distribution of answers,
leading to a uniform frequency distribution. 

We use entropy as a metric of consistency. The uniform distribution (even
distributed answers) leads to maximum entropy, while full agreement on 1
partition leads to lowest entropy. Then, a consistent mental model is one that
produces a low entropy distribution on the answers to the DGIC questions.

\textit{Using Entropy Adequately.} Using entropy introduces a challenge in
comparing consistencies because different \KR s contain different number of
partitions. A lower entropy may be caused for the
following reasons that would not correspond to high consistency: (I1) 
the MMs have vastly different numbers of partitions. The fewer partitions
a MM has, the lower its entropy can be, so results would be skewed in favor of
the smaller MM. (I2) there are fewer respondents than the number of partitions
for a question. Then, low entropy may be misleading because at least one
partition can never be chosen. We made sure none of these reasons apply to our
setting. I1 does not apply because the number of MM partitions is roughly the
same: 5W1H+R and GCS both have 7 partitions, and Datahub has 6. I2 does not
apply because in all cases, we made sure there were more participants than
number of partitions.

That said, we normalize the entropy to help interpret the
results and address I1 above.  This allows us to measure the extent to which a
distribution differs from uniform on a scale from 0 to 1, where 0 indicates that
the distribution is farthest from uniform, and 1 indicates that the distribution
is exactly uniform. Therefore, we can compare the normalized entropies of two
frequency distributions to conclude that the one with lower normalized entropy
is farther from uniform (more consistent) than the other. 

Finally, we also analyze \textsc{None} response frequency. A \KR\ that causes
participants to choose \textsc{None} more often is less comprehensible because
it makes it less likely that data users will agree on a partition for
storing/retrieving a MI.

\noindent\textbf{Ease-of-Use:} a \KR's ease-of-use for a data question is
measured asking data users to score the difficulty, which we do using the
standard Likert scale (a 1-5 difficulty rating, with 1 being "Very Easy", and 5
being "Very Difficult"). 

\subsubsection{Participant Recruitment}
\label{subsubsec:partrec}

We use Prolific for running the user study because it respects users' privacy
better, being based in UK and hence subject to GDPR and giving users the option
to remove their data. Further, Prolific's completion rate is higher than
that of other 
crowd platforms. We use a combination of
Prolific filters to make sure that only participants with formal education can
participate, as we expect these are the ones in positions of data engineering,
business analysts, and other data users. Furthermore, we make sure English is
their first language because the study is written in English.

In addition to the Prolific filters, we run a thorough pre-screening survey. To
identify real data users, we ask participants to rate their familiarity with our
questions. We only ask them to classify questions they rather with high
familiarity, thus avoiding confounding effects related to lack of understanding,
lack of context, or lack of skill to answer a question. We
also ask participants which catalogs they are familiar with and we do not show a
participant the MM from a catalog they are familiar with.

\subsubsection{Threats To Validity}
\label{subsec:threats}

We conducted several pilot studies with database researchers
and non-technical people
alike to hone the survey language and identify threats to validity. Based on our
findings we implemented the following measures during the study design to avoid
such threats: 

\noindent\textit{Ordering and Learning Effects}: We use a between-subjects
study, which complicates our statistical analysis and data collection, but
avoids learning effects of answering questions across \KR s because each
participant is shown only one \KR.

\noindent\textit{Selection Effects}: We use the prescreening questions and
Prolific filters to select a homogeneous participant pool with data management
familiarity. We assign \KR s to participants in the pool randomly, and they only
answer questions they understand.

\noindent\textit{Experimenter Bias}: We mitigate the effects of experimenter
bias by generating all \KR s using the same procedure (Section
\ref{subsec:targets}), and presenting these \KR s to participants with the same
instructions on how to use each. Because the study is an online
survey, participants are not further exposed to our bias.

\noindent\textit{Reactivity Effects}: We ask participants to rate the difficulty
of categorizing data questions. We mitigate the potential Hawthorne
effect~\cite{hawthorne} by wording the questions carefully (after several pilot
studies) and by explaining in the survey introduction that the survey is intended
to evaluate the \KR s and not the participants' ability. 

\subsection{Study Results}
\label{subsec:consistency}

We present the consistency and ease-of-use results. For all results, we say 5W1H+R MM is \emph{Better} if it outperforms both MMs, \emph{Inconclusive} if it outperforms only one, and \emph{Worse} if it outperforms neither.

\subsubsection{Consistency} The 5W1H+R \KR\ is more consistent than the others:
it achieves lower entropy and participants choose \emph{None} significantly less
frequently than in other MM. 

\input{tables/entropyresults.tex}

\input{tables/infefficiencyresults.tex}

\input{tables/noneresults.tex}

\mypar{Entropy} The entropy results of
Table~\ref{all_entropies} show that the 5W1H+R \KR\ has lower entropy than
others in 15 questions, and is worse in only 2/27. When normalizing the entropy
per number of concepts in the mental model, the results hold.
Table~\ref{all_efficiencies} show the 5W1H+R \KR\ has lower normalized entropy
than others in 16 questions, and that it is worse in only 2/27. Furthermore, it
has a lower average normalized entropy overall. 

\mypar{Choosing None} \textsc{None} answers indicate a
weakness of a \KR\: the greater their frequency, the less likely it is
that \KR\ users will agree on a partition \textit{from the MM} for
storing/retrieving a MI. Table \ref{nonestable} shows that participants using
5W1H+R chose this option with the lowest proportion out of all responses, and
with the least absolute amount; \emph{less than half as frequently as in other}
\KR s. In 18 out of 27 questions, the lowest proportion of participants chose
it compared to other \KR s. Finally, \textsc{None} was the most selected answer
out of the \KR\ partitions for only 7 questions when using the 5W1H+R \KR, as
opposed to 19 and 16 for GCS and Datahub, respectively.

\textbf{Lower entropy and lower use of \textsc{None} demonstrate the 5W1H+R \KR\
has higher consistency, and is therefore more comprehensible.}

\subsubsection{Consistency Results in Depth} 
\label{subsec:indepth}

In this section, we explain the results in greater depth by zooming in on
details on the frequency distributions for 2 data questions where 5W1H+R
outperforms others, and 1 data question where it does not.
\suggestion{R1O1}{One potential threat to validity that may explain the success of 5W1H+R MM compared to other MMs is the similarity in keywords between 5W1H+R MM partitions and the questions we used in our experiments (e.g., a question that begins with What is a what-profile, Where is where-profile, etc.). While we do not observe this similarity (as we will see below), we also argue that this is not an actual threat to validity. Rather, it is another property of the 5W1H+R MM that can make it more comprehensible compared to other MMs.}

\begin{figure*}[!htb]
   \begin{minipage}{0.32\textwidth}
     \centering
     \includegraphics[width=.9\linewidth]{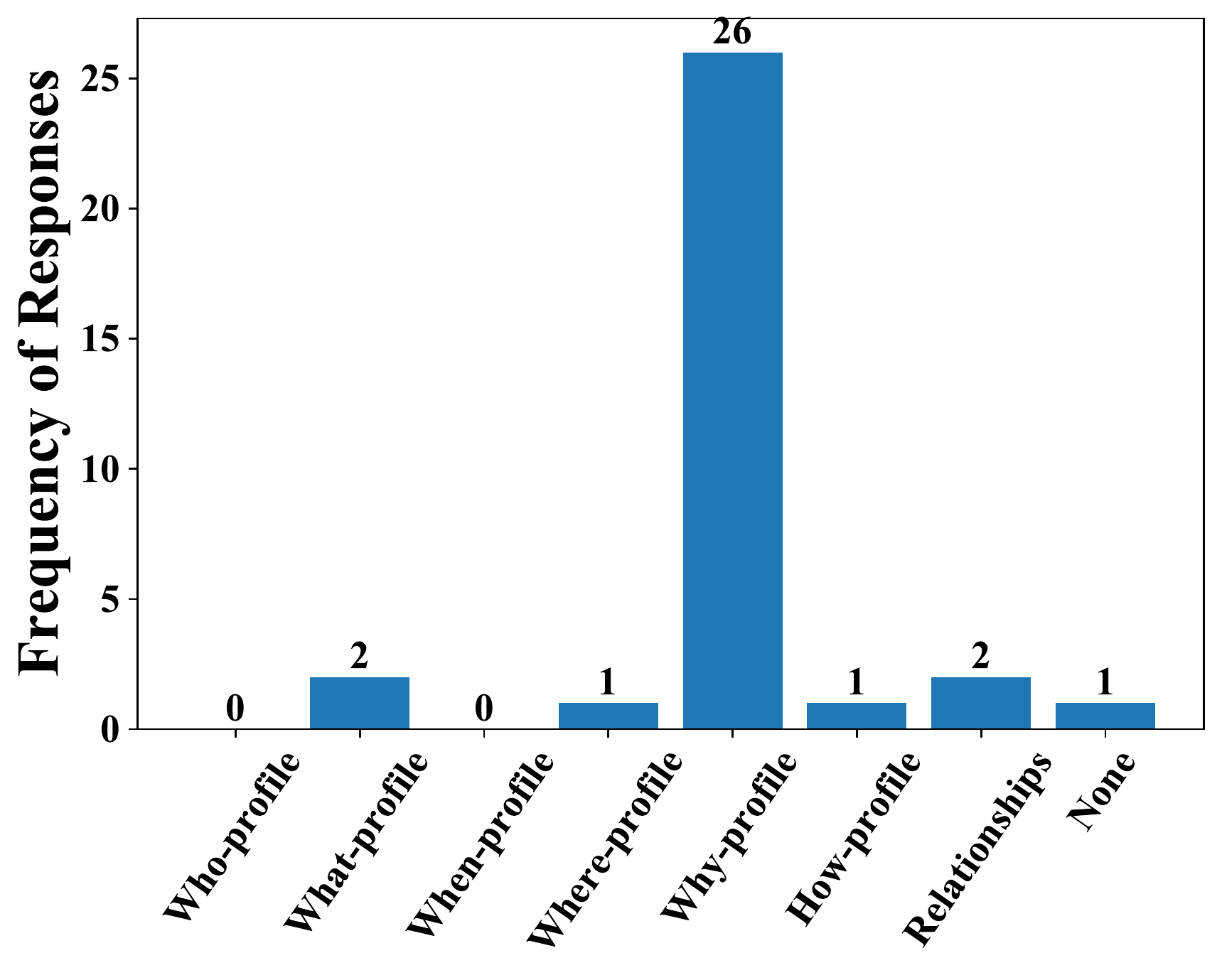}
   \end{minipage}\hfill
   \begin{minipage}{0.32\textwidth}
     \centering
     \includegraphics[width=.9\linewidth]{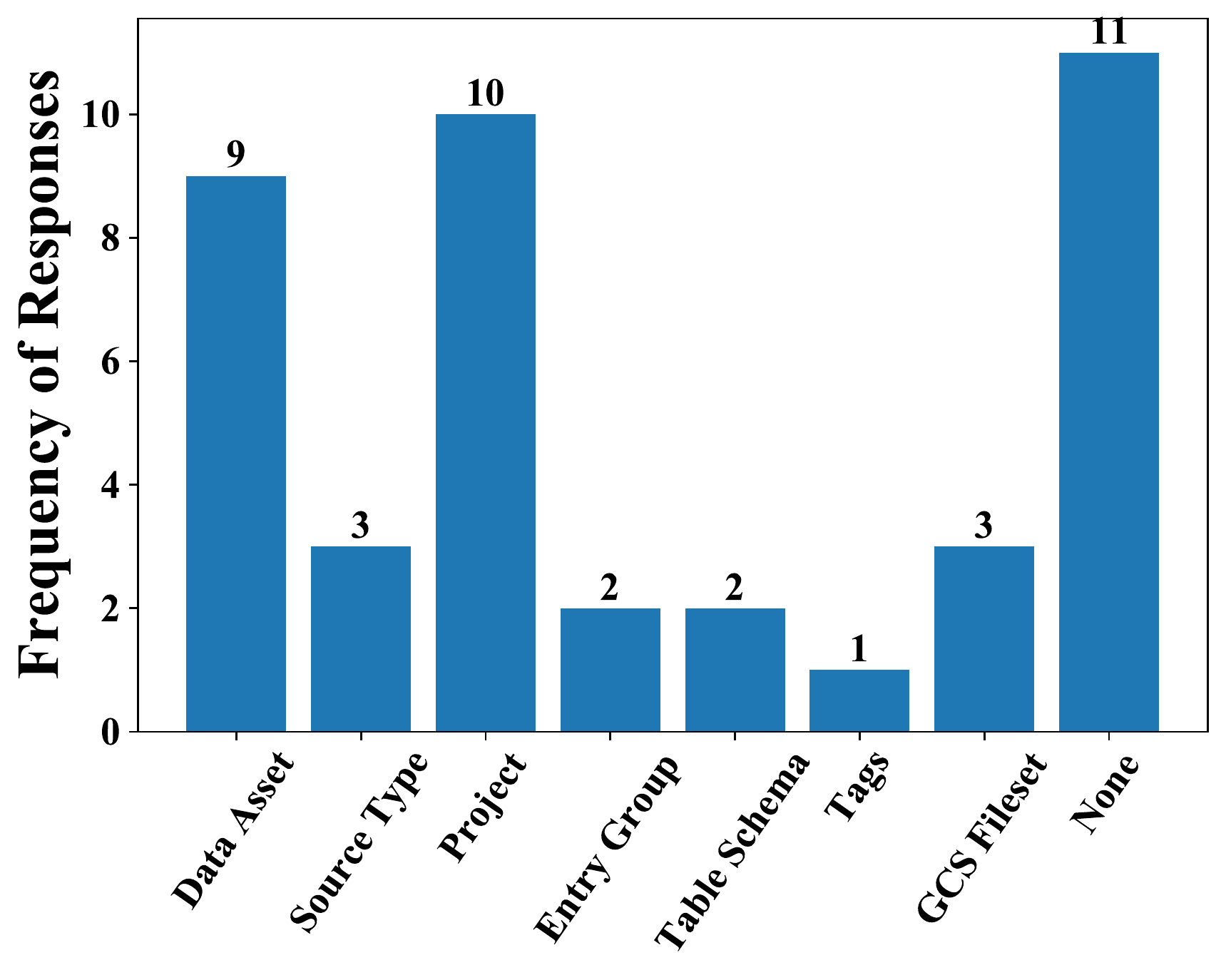}
   \end{minipage}
   \begin{minipage}{0.32\textwidth}
     \centering
     \includegraphics[width=.9\linewidth]{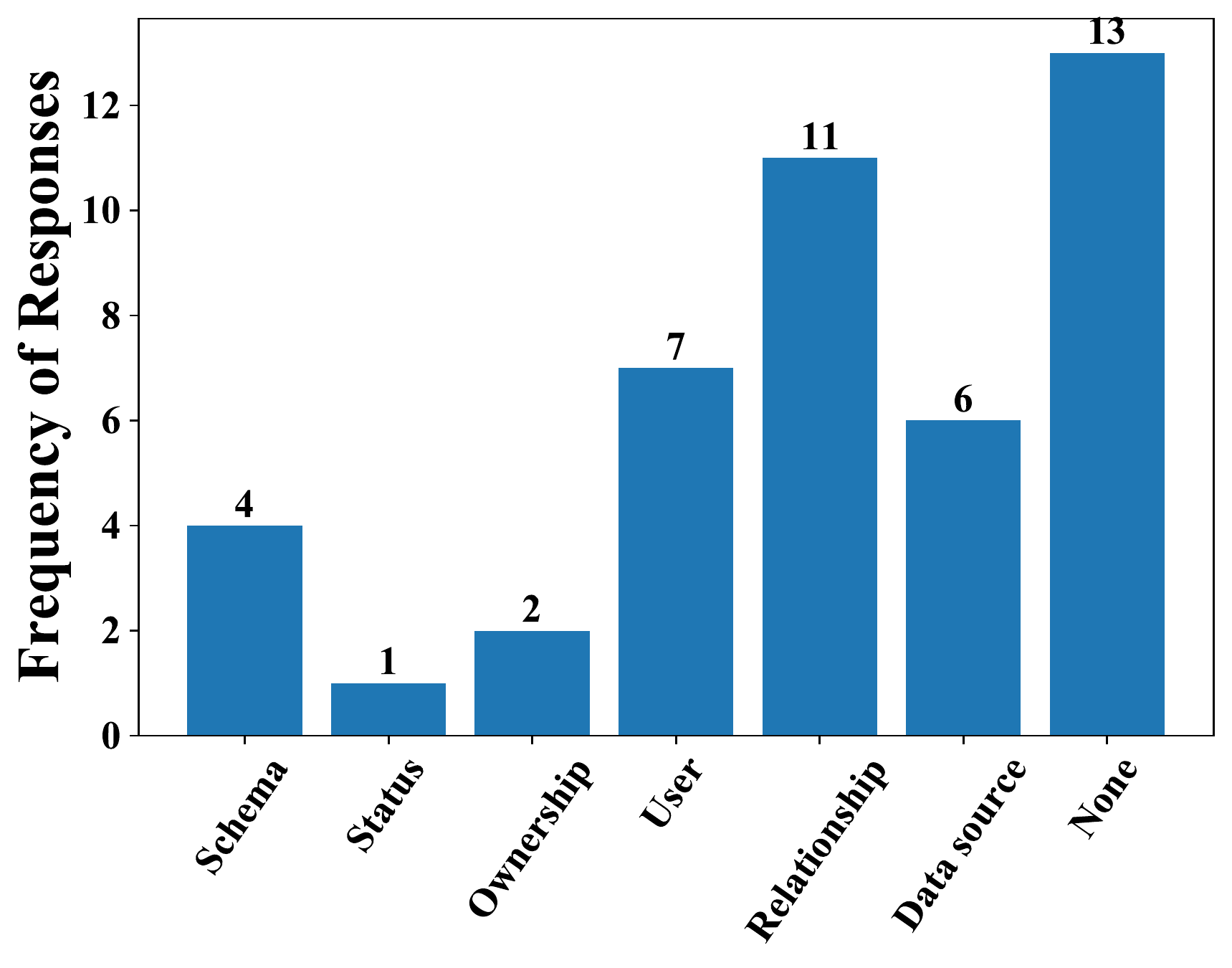}
   \end{minipage}
   \caption{Q1 Frequency Distributions for 5W1H+R (Left), GCS (Center) and Datahub (Right)}  \label{fig:Q1freqs}
\end{figure*}

\begin{figure*}[!htb]
   \begin{minipage}{0.32\textwidth}
     \centering
     \includegraphics[width=.9\linewidth]{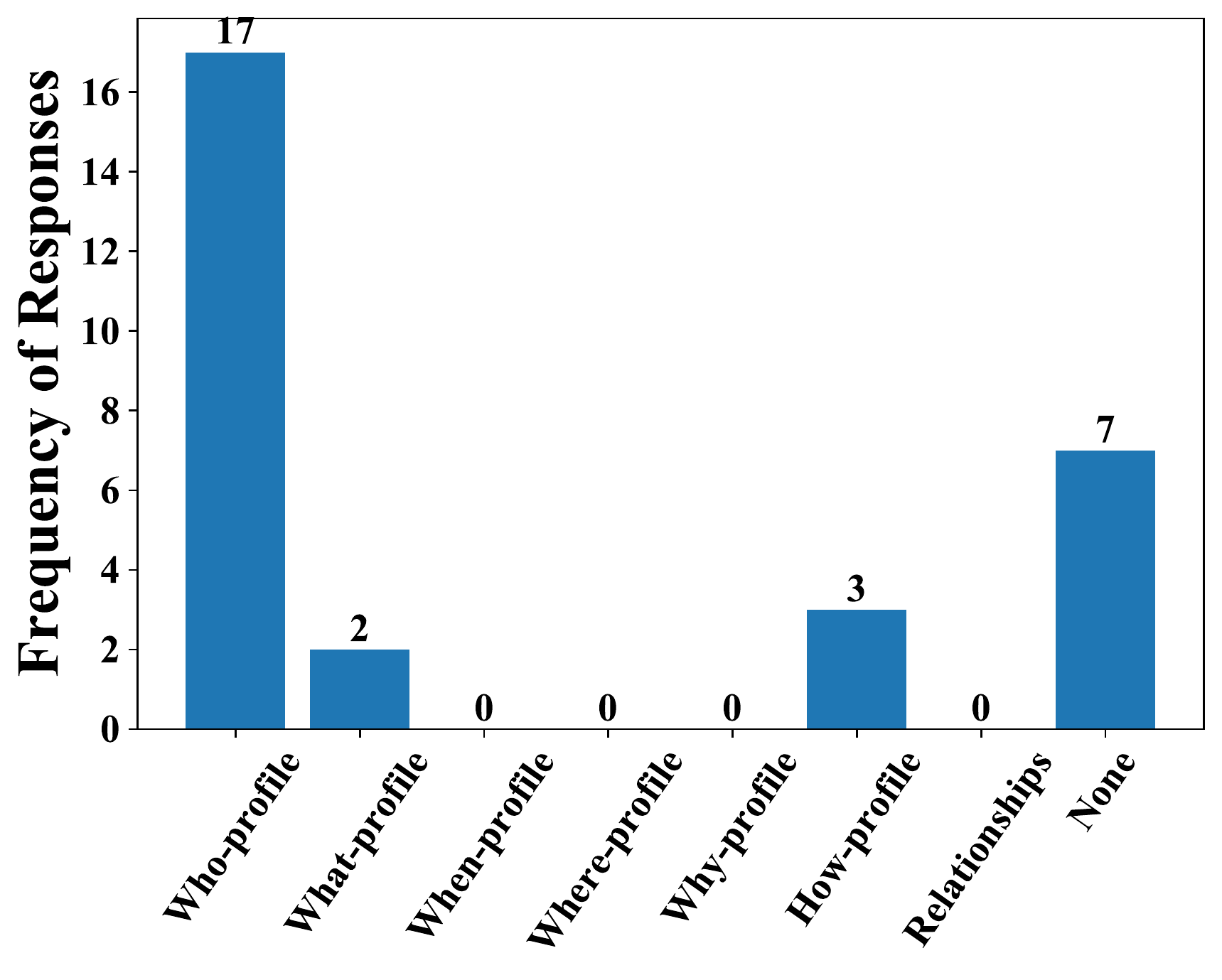}
   \end{minipage}\hfill
   \begin{minipage}{0.32\textwidth}
     \centering
     \includegraphics[width=.9\linewidth]{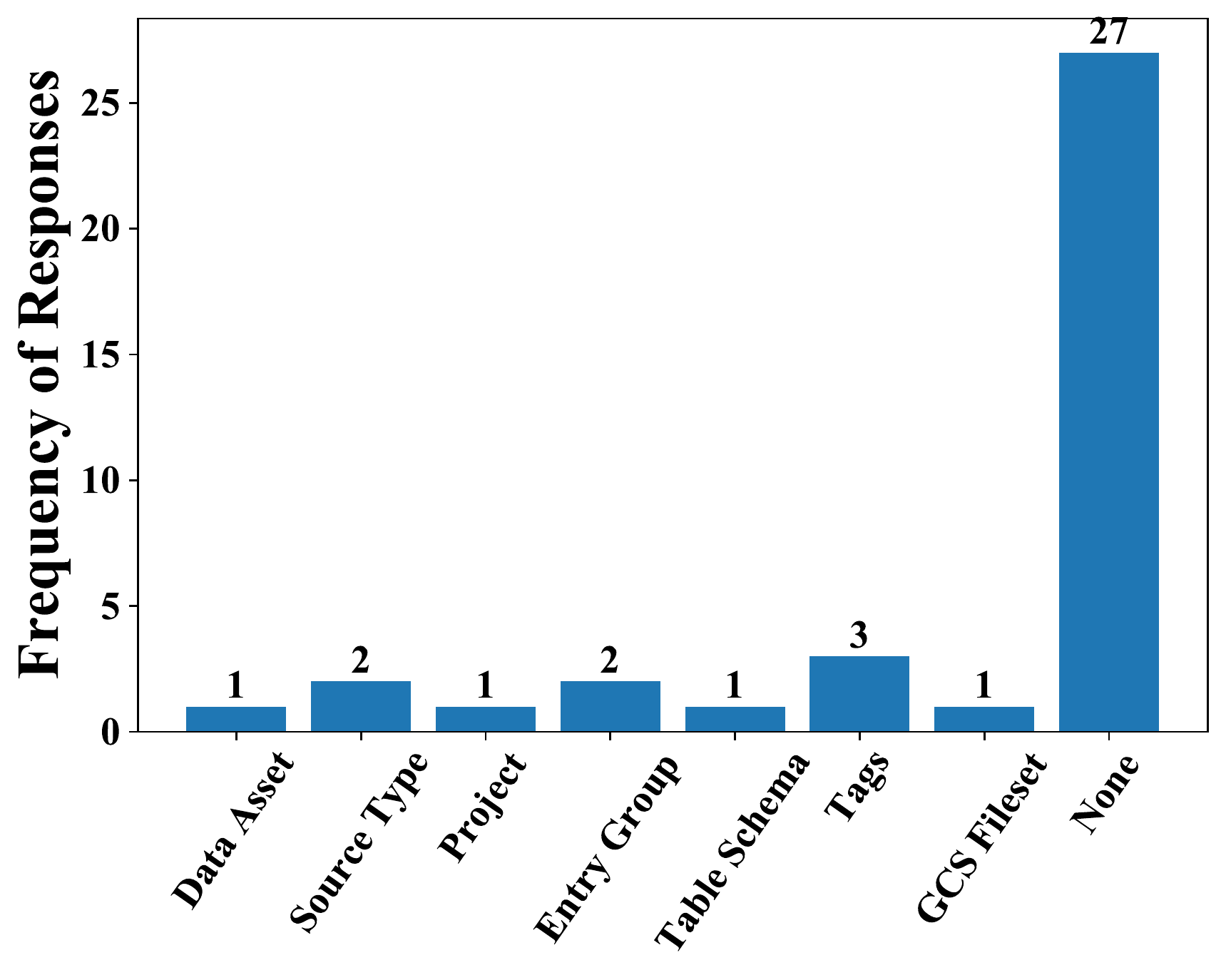}
   \end{minipage}
   \begin{minipage}{0.32\textwidth}
     \centering
     \includegraphics[width=.9\linewidth]{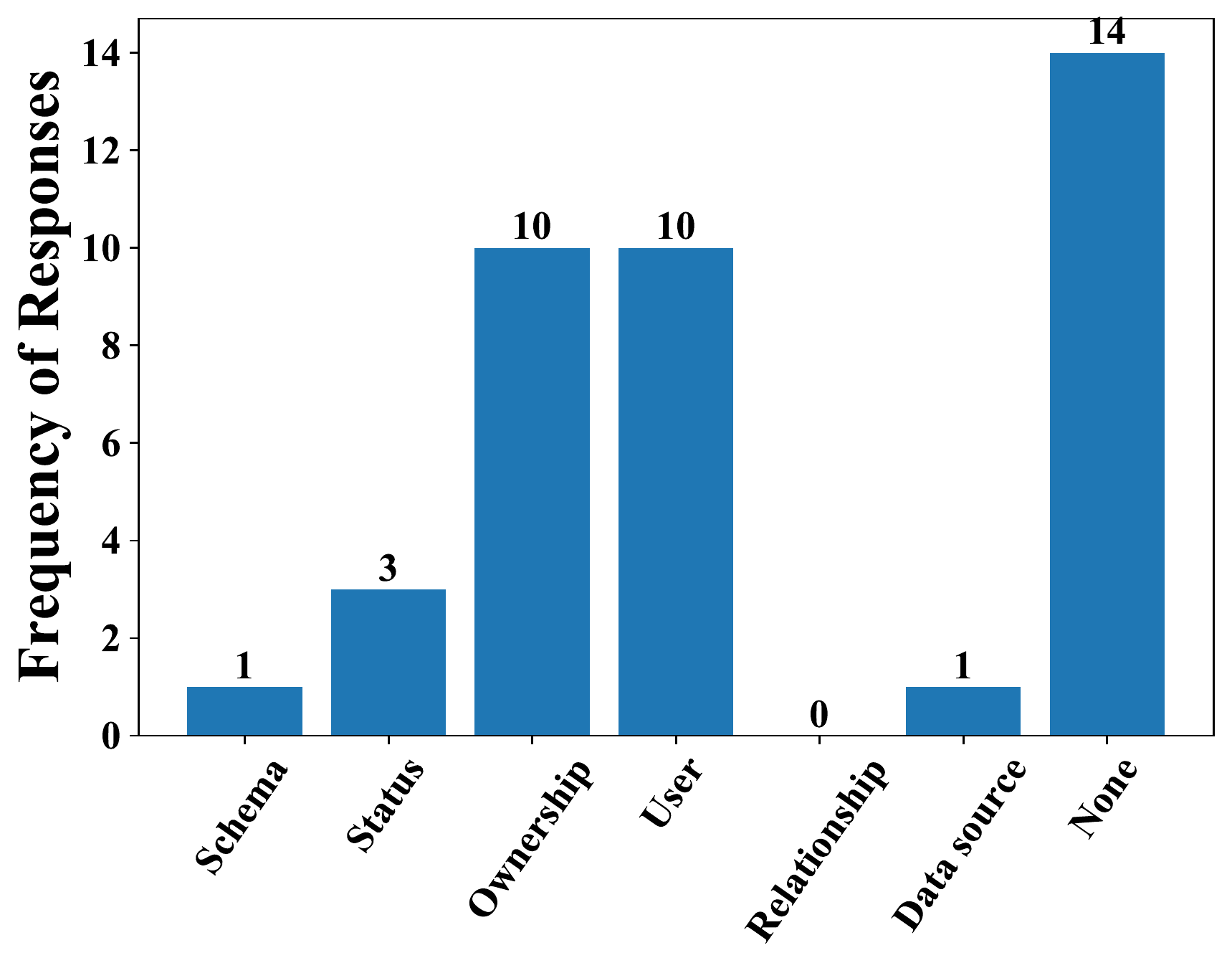}
   \end{minipage}
   \caption{Q8 Frequency Distributions for 5W1H+R (Left), GCS (Center) and Datahub (Right)} \label{fig:Q8freqs}
\end{figure*}

\begin{figure*}[!htb]
   \begin{minipage}{0.32\textwidth}
     \centering
     \includegraphics[width=.9\linewidth]{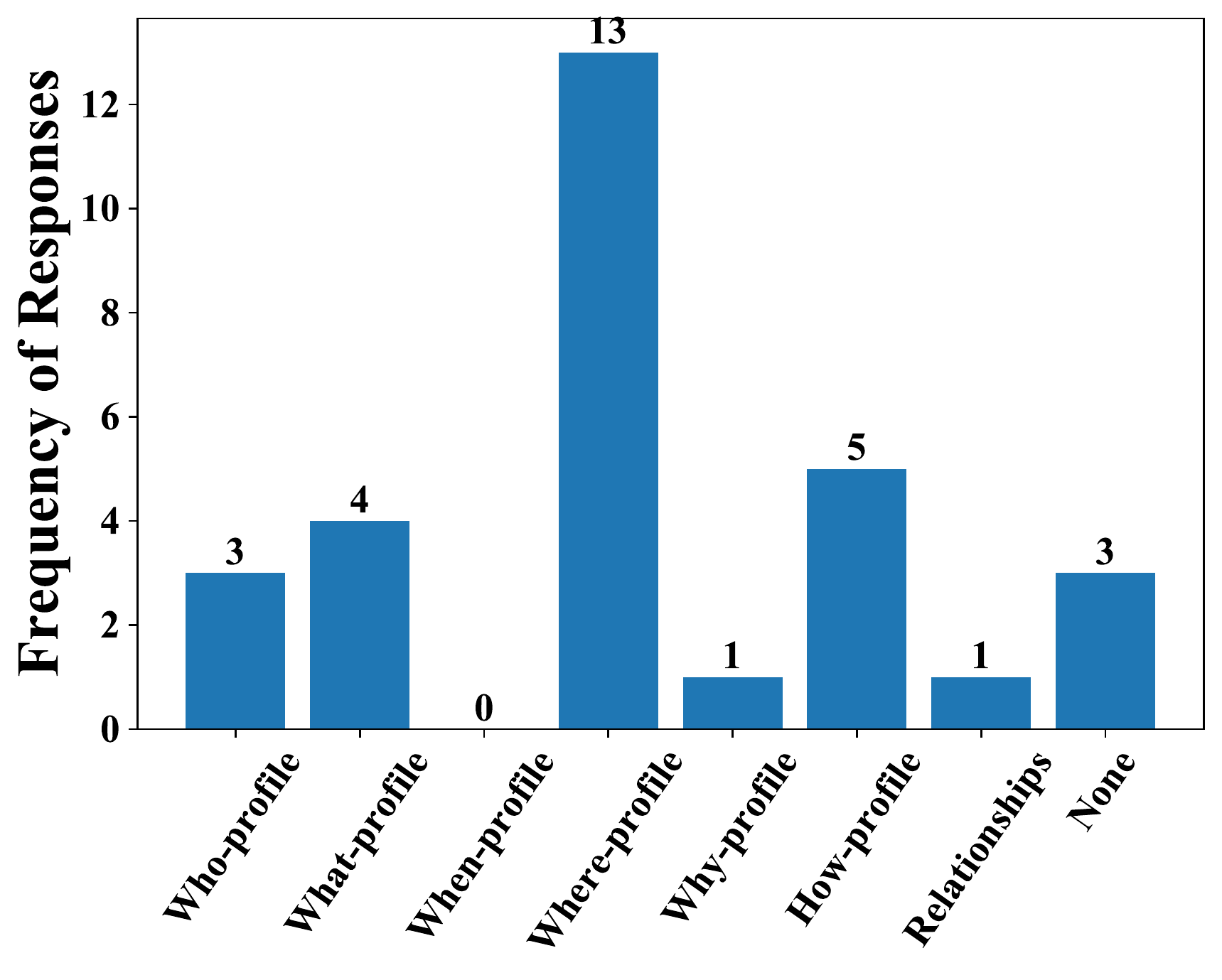}
   \end{minipage}\hfill
   \begin{minipage}{0.32\textwidth}
     \centering
     \includegraphics[width=.9\linewidth]{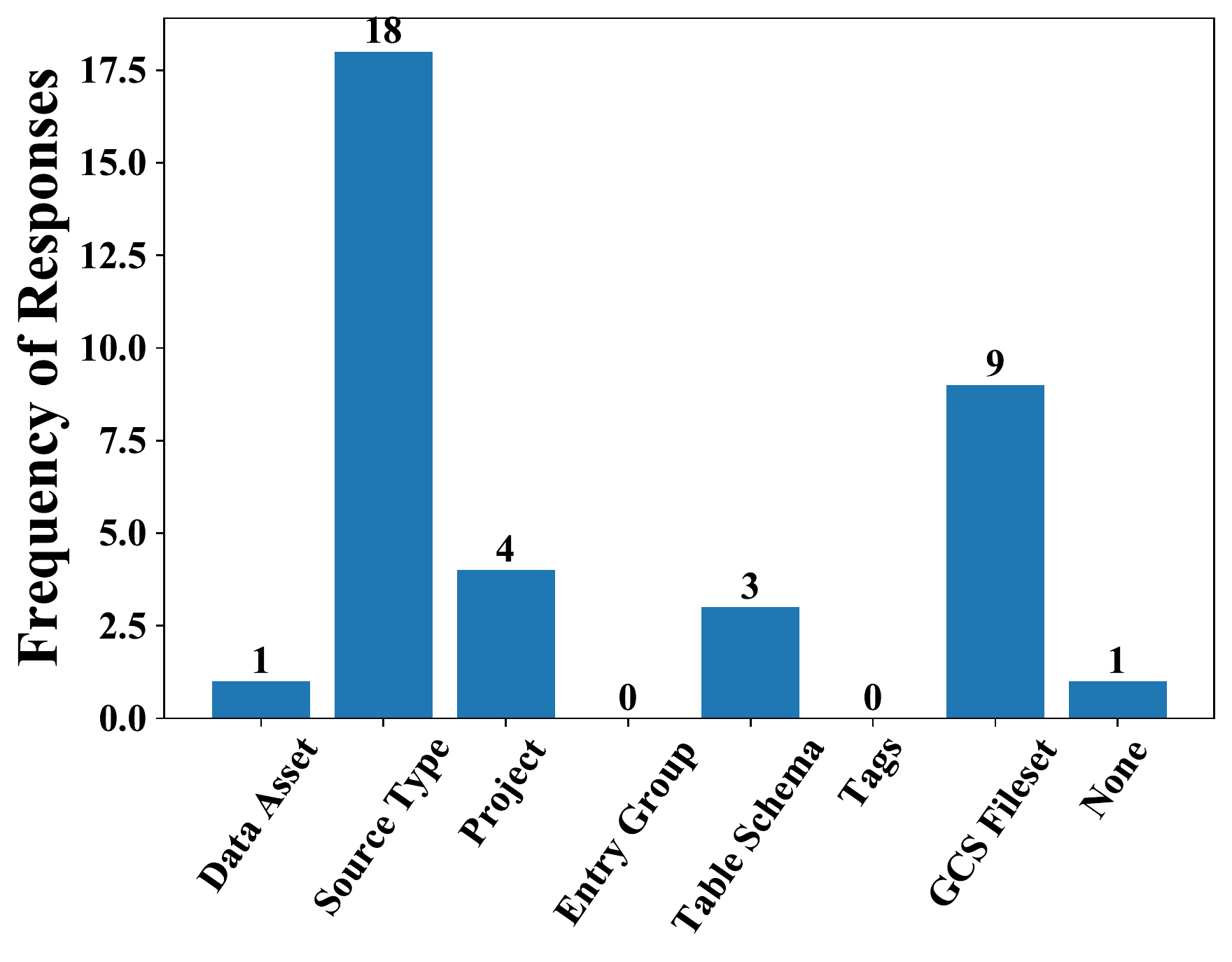}
   \end{minipage}
   \begin{minipage}{0.32\textwidth}
     \centering
     \includegraphics[width=.9\linewidth]{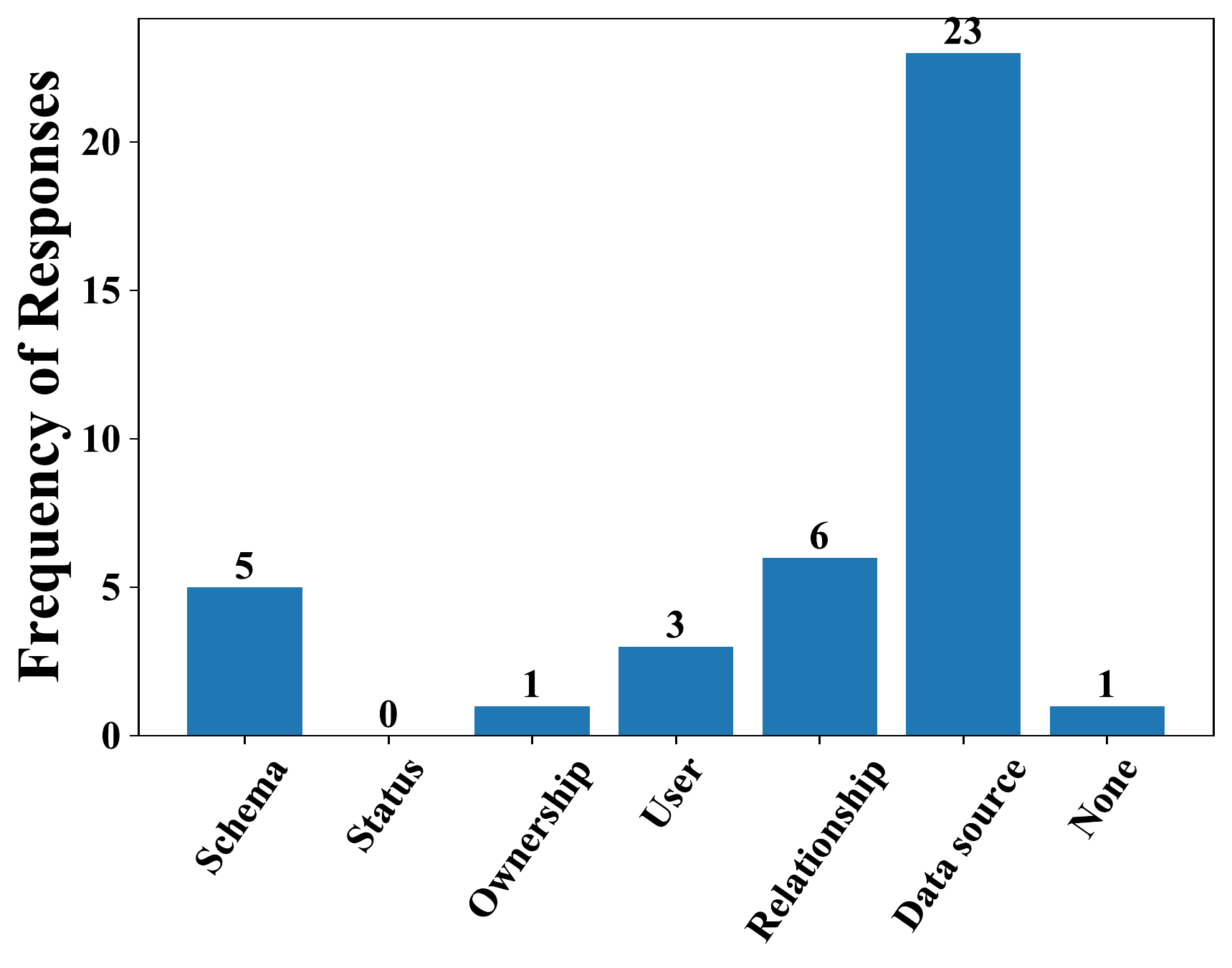}
   \end{minipage}
   \caption{Q25 Frequency Distributions for 5W1H+R (Left), GCS (Center) and Datahub (Right)} \label{fig:Q25freqs}
\end{figure*}

\input{tables/ratingresults.tex}

\mypar{Q1-For what purpose was the dataset created?} From
Table~\ref{all_entropies}, we see that the entropy of the 5W1H+R \KR\ is 1.06,
whereas it is 2.4 for GCS, and 2.27 for Datahub, a difference of at least 1.21.
It is obvious from the frequency distributions in \F\ref{fig:Q1freqs} why 5W1H+R
outperforms GCS and Datahub: most 5W1H+R participants chose Why-profile to
answer this question, and very few others chose any other answer. In contrast,
GCS and Datahub participants were more divided on this question.

We interpret the underlying reason for 5W1H+R's better consistency in this case
to be that it is more comprehensive compared to other \KR s, because there is no
partition in either GCS or Datahub that explicitly captures the purpose of a
dataset, even though this information is commonly required in practice when
machine learning engineers need to understand the origin of a training dataset
\cite{gebru2018datasheets}, or when scientists need to understand whether data
they already have can be used for a new purpose \cite{dataworkref}.  On the
other hand, the definition of "Why-profile" clearly includes this question.  The
higher number of \textsc{None}s in the GCS and Datahub \KR s further supports
this interpretation: the 5W1H+R \KR\ has only 1 \textsc{None} answer, but GCS
and Datahub have 11 and 13, respectively.

\mypar{Q8-What is the reputation of the creator of a dataset?} From
Table~\ref{all_entropies}, we see that the entropy of the 5W1H+R \KR\ is 0.99,
whereas it is 2.66 for GCS and 1.8 for Datahub, a difference of at least 0.81.
For this question, note that the Datahub frequency distribution in
\F\ref{fig:Q8freqs} has two peaks: one for the "User" partition, and one for the
"Ownership" partition, whereas the most popular partition for the 5W1H+R \KR\ is
"Who-profile".  Our interpretation of this is that participants deduced that
reputation could be inferred from information about a user of a dataset.
However, because there were two partitions for user-related metadata,
participants could not agree on where this metadata would be found.  This
example illustrates how a catalog user who is trying to determine a dataset
creator's reputation can be confused about whether to find the creator's most
up-to-date details in "Users", or "Ownership" if they have to rely on the
Datahub \KR.  Therefore, our interpretation of the underlying reason for
5W1H+R's better consistency in this case is that it has a single partition for
metadata about data users (including creators of datasets).  Therefore, there is
no disagreement on where to find data-user-related metadata.

5W1H+R outperforms GCS for the same reason as before: there is no
clear consensus from participants who chose a partition from GCS
(\F\ref{fig:Q8freqs}), and \textsc{None} was the most popular answer. This
indicates there is no partition that captures the reputation of the creator of a
dataset in GCS.

\mypar{Q25-What is the format of the dataset, and what type of repository is the
dataset located in?} We complement our in-depth analysis with a question where
the 5W1H+R \KR\ does worse than the others. From Table~\ref{all_entropies}, we
see that the entropy of the 5W1H+R \KR\ is 2.07, whereas it is 1.81 in GCS, and
1.67 in Datahub, a difference of only 0.26.  From the frequency
distributions in \F\ref{fig:Q25freqs}, we see that 5W1H+R underperforms in this
case: participants were more divided when using 5W1H+R than in the GCS case,
where 18 participants chose the "Source Type" partition, and very few
participants chose any other partition, or the Datahub case, where 23
participants chose the "Data source" partition, and very few participants chose
any other answer. (But "Where-profile" still appears to have been chosen over
the other partitions).

We interpret the underlying reason for 5W1H+R's worse consistency to be that the
other \KR s have partitions whose definitions explicitly capture the answer by
using more concrete language: in \F\ref{fig:gcsMM} and \F\ref{fig:datahubMM}, the definitions of "Source Type" and "Data source" clearly answers this
question, and the term "repository" appears in these definitions, along with
examples of repository formats.  In contrast, the "Where-profile" partition
definition does not explicitly say that a Where-profile records metadata about
repositories and their formats.  This is a case where the 5W1H+R \KR\ suffers
because it does not provide any entities for describing metadata beyond the
profiles. We discuss how more specific entities from ontologies could be used to
complement the 5W1H+R \KR\ in Section \ref{sec:discussion}.

\subsubsection{Results: Ease of Use}
\label{subsec:ease}

Table~\ref{ratingstable} shows that the 5W1H+R \KR\ has the same or lower median
difficulty than the others for 25 out of the 27 questions. Further, using a
Mann-Whitney U Test, we find that there is a significant difference between the
5W1H+R median difficulty rating and the median difficulty rating of at least one
of the other \KR s for 16 of the 25 questions.  This result suggests that for
most questions, \textbf{the 5W1H+R \KR\ is easier to use compared to other \KR
s}. In summary, this means participants had an easier time deciding where to
classify a question MI when using 5W1H+R than when using other MM.

\mypar{DGIC Tasks with 5W1H+R} One may wonder if the above consistency and
ease-of-use results apply similarly for each of the DGIC question categories.
Stratifying the results of Tables \ref{all_entropies} and \ref{ratingstable} by
DGIC category generates similar consistency and ease-of-use results
within each category. This indicates the 5W1H+R MM reduces the effort required
for DGIC tasks and is further indication of its comprehensiveness.

%% file: tables/entropyresults.tex
\begin{table*}[!htbp]
\centering
\resizebox{\textwidth}{!}{%
\begin{tabular}{@{}lrrrrrrrrrrrrrrrrrrrrrrrrrrrl@{}}
\toprule
\multicolumn{16}{c|}{Better} &
  \multicolumn{10}{c|}{Inconclusive} &
  \multicolumn{2}{c|}{Worse} &
  Average \\ \midrule
MM &
  Q1 &
  Q2 &
  Q3 &
  Q4 &
  Q8 &
  Q9 &
  Q10 &
  Q13 &
  Q15 &
  Q17 &
  Q18 &
  Q19 &
  Q22 &
  Q23 &
  Q27 &
  Q5 &
  Q6 &
  Q7 &
  Q11 &
  Q12 &
  Q14 &
  Q16 &
  Q20 &
  Q21 &
  Q24 &
  Q25 &
  Q26 &
   \\ \midrule
5W1H &
  \textbf{1.06} &
  \textbf{1.34} &
  \textbf{0.81} &
  \textbf{0.93} &
  \textbf{0.99} &
  \textbf{1.83} &
  \textbf{1.73} &
  \textbf{1.9} &
  \textbf{1.94} &
  \textbf{1.83} &
  \textbf{2} &
  \textbf{2.37} &
  \textbf{1.6} &
  \textbf{0.61} &
  \textbf{1.58} &
  1.52 &
  2.42 &
  1.97 &
  1.81 &
  2.25 &
  2.06 &
  2.09 &
  1.46 &
  1.13 &
  1.84 &
  2.07 &
  2.72 &
  \textbf{1.698} \\
GCS &
  2.4 &
  2.72 &
  2.6 &
  2.52 &
  2.66 &
  2.68 &
  2.48 &
  2.67 &
  2.32 &
  2.4 &
  2.49 &
  2.61 &
  2.05 &
  1.01 &
  2.28 &
  2.12 &
  2.44 &
  \textbf{1.78} &
  2.35 &
  2.36 &
  2.49 &
  2.48 &
  \textbf{1.44} &
  \textbf{0.68} &
  2.02 &
  1.81 &
  1.88 &
  2.212 \\
Datahub &
  2.27 &
  2.15 &
  1.65 &
  1.34 &
  1.8 &
  2.36 &
  2.18 &
  2.52 &
  2.01 &
  2.25 &
  2.37 &
  2.38 &
  1.91 &
  1.23 &
  1.66 &
  \textbf{1.41} &
  \textbf{2.02} &
  1.92 &
  \textbf{1.73} &
  \textbf{2.03} &
  \textbf{1.63} &
  \textbf{1.87} &
  2.06 &
  1.64 &
  \textbf{1.65} &
  \textbf{1.67} &
  \textbf{1.81} &
  1.903 \\ \bottomrule
\end{tabular}%
}
\caption{Entropies per \KR\ per Question}
\label{all_entropies}
\end{table*}

%% file: tables/infefficiencyresults.tex
\begin{table*}[!htbp]
\resizebox{\textwidth}{!}{%
\begin{tabular}{@{}lrrrrrrrrrrrrrrrrrrrrrrrrrrrl@{}}
\toprule
\multicolumn{17}{c|}{Better} &
  \multicolumn{9}{c|}{Inconclusive} &
  \multicolumn{2}{c|}{Worse} &
  Average \\ \midrule
\textbf{MM} &
  Q1 &
  Q2 &
  Q3 &
  Q4 &
  Q8 &
  Q9 &
  Q10 &
  Q11 &
  Q13 &
  Q15 &
  Q17 &
  Q18 &
  Q19 &
  Q22 &
  Q23 &
  Q27 &
  Q5 &
  Q6 &
  Q7 &
  Q12 &
  Q14 &
  Q16 &
  Q20 &
  Q21 &
  Q24 &
  Q25 &
  Q26 &
   \\ \midrule
5W1H &
  \textbf{0.37} &
  \textbf{0.47} &
  \textbf{0.29} &
  \textbf{0.33} &
  \textbf{0.35} &
  \textbf{0.65} &
  \textbf{0.62} &
  \textbf{0.64} &
  \textbf{0.67} &
  \textbf{0.69} &
  \textbf{0.65} &
  \textbf{0.71} &
  \textbf{0.84} &
  \textbf{0.57} &
  \textbf{0.22} &
  \textbf{0.56} &
  \textbf{0.54} &
  0.86 &
  0.7 &
  0.8 &
  0.73 &
  0.74 &
  0.52 &
  0.4 &
  0.65 &
  0.73 &
  0.96 &
  \textbf{0.603} \\
GCS &
  0.85 &
  0.97 &
  0.92 &
  0.89 &
  0.95 &
  0.95 &
  0.88 &
  0.83 &
  0.95 &
  0.82 &
  0.85 &
  0.88 &
  0.93 &
  0.73 &
  0.36 &
  0.81 &
  0.75 &
  0.87 &
  \textbf{0.63} &
  0.84 &
  0.88 &
  0.88 &
  \textbf{0.51} &
  \textbf{0.24} &
  0.72 &
  \textbf{0.64} &
  \textbf{0.67} &
  0.785 \\
Datahub &
  0.86 &
  0.82 &
  0.63 &
  0.51 &
  0.68 &
  0.9 &
  0.83 &
  0.66 &
  0.96 &
  0.77 &
  0.86 &
  0.9 &
  0.91 &
  0.73 &
  0.47 &
  0.63 &
  \textbf{0.54} &
  \textbf{0.77} &
  0.73 &
  \textbf{0.77} &
  \textbf{0.62} &
  \textbf{0.71} &
  0.78 &
  0.63 &
  \textbf{0.63} &
  \textbf{0.64} &
  0.69 &
  0.727 \\ \bottomrule
\end{tabular}%
}
\caption{Normalized Entropies per \KR\ per Question}
\label{all_efficiencies}
\end{table*}

%% file: tables/noneresults.tex
\begin{table*}[!htbp]
\centering
\resizebox{\textwidth}{!}{%
\begin{tabular}{@{}llllrllrlrlllllllllllllrlrllrll@{}}
\toprule
MM &
  Q1 &
  Q2 &
  Q3 &
  Q4 &
  Q5 &
  Q6 &
  Q7 &
  Q8 &
  Q9 &
  Q10 &
  Q11 &
  Q12 &
  Q13 &
  Q14 &
  Q15 &
  Q16 &
  Q17 &
  Q18 &
  Q19 &
  Q20 &
  Q21 &
  Q22 &
  Q23 &
  Q24 &
  Q25 &
  Q26 &
  Q27 &
  \textsc{Nones} &
  Responses &
  Proportions \\ \midrule
5W1H &
  \textbf{0.03} &
  \textbf{0.39} &
  \textbf{0.03} &
  \textbf{0.06} &
  0.12 &
  \textbf{0.23*} &
  0.14 &
  \textbf{0.24} &
  0.23 &
  0.30 &
  0.56* &
  0.39* &
  0.15 &
  \textbf{0.03} &
  \textbf{0.19} &
  \textbf{0.34*} &
  \textbf{0.23} &
  \textbf{0.27} &
  \textbf{0.29*} &
  \textbf{0.03} &
  \textbf{0.06} &
  \textbf{0.16} &
  \textbf{0.06} &
  0.35* &
  0.1 &
  \textbf{0.27*} &
  \textbf{0.06} &
  165 &
  832 &
  0.199 \\
GCS &
  0.27* &
  0.42* &
  0.34* &
  0.2 &
  0.33* &
  0.46* &
  0.15 &
  0.71* &
  \textbf{0.12} &
  \textbf{0.27*} &
  0.68* &
  0.51* &
  0.20* &
  0.24* &
  0.30* &
  0.51* &
  0.42* &
  0.47* &
  0.51* &
  0.12 &
  0.12 &
  0.42* &
  0.2 &
  \textbf{0.20} &
  \textbf{0.03} &
  0.35* &
  0.25* &
  340 &
  1033 &
  0.329 \\
Datahub &
  0.30* &
  0.61* &
  0.11 &
  0.24 &
  \textbf{0.09} &
  0.48* &
  \textbf{0.08} &
  0.36* &
  0.19 &
  0.43* &
  \textbf{0.22} &
  \textbf{0.36*} &
  0.22* &
  0.21 &
  0.37* &
  0.37* &
  0.33* &
  0.42* &
  0.43* &
  0.40* &
  0.44* &
  0.44* &
  0.09 &
  0.34 &
  \textbf{0.03} &
  0.36* &
  0.12 &
  331 &
  1109 &
  0.298 \\ \bottomrule
\end{tabular}%
}
\caption{\label{nonestable} Proportions and Numbers of \textsc{None} Responses Per \KR\ Across Questions And Users: * \textsc{None} was the most selected answer}
\end{table*}

%% file: tables/ratingresults.tex
\begin{table*}[!htbp]
\resizebox{\textwidth}{!}{%
\begin{tabular}{@{}lrrrrrrrrrrrrrrrrrrrrrrrrrrr@{}}
\toprule
\multicolumn{9}{c|}{Better} &
  \multicolumn{17}{c|}{Inconclusive} &
  \multicolumn{2}{c|}{Worse} \\ \midrule
MM &
  Q1 &
  Q2 &
  Q3 &
  Q4 &
  Q9 &
  Q18 &
  Q20 &
  Q27 &
  Q5 &
  Q6 &
  Q7 &
  Q8 &
  Q10 &
  Q12 &
  Q13 &
  Q14 &
  Q15 &
  Q16 &
  Q17 &
  Q19 &
  Q21 &
  Q22 &
  Q23 &
  Q24 &
  Q26 &
  Q11 &
  Q25 \\ \midrule
5W1H &
  \textbf{2**} &
  \textbf{2**} &
  \textbf{2**} &
  \textbf{2**} &
  \textbf{3*} &
  \textbf{3*} &
  \textbf{2*} &
  \textbf{2*} &
  3 &
  \textbf{3*} &
  3* &
  3 &
  3 &
  \textbf{3} &
  3** &
  \textbf{3**} &
  3 &
  3 &
  3 &
  \textbf{3*} &
  \textbf{2*} &
  3** &
  2 &
  3* &
  \textbf{3} &
  4 &
  3 \\
GCS &
  3 &
  3 &
  3 &
  3 &
  3.5 &
  4 &
  3 &
  3.5 &
  3 &
  4 &
  3 &
  3 &
  3 &
  4 &
  3 &
  3.5 &
  3 &
  3 &
  3 &
  \textbf{3} &
  \textbf{2} &
  3 &
  2 &
  3 &
  4 &
  4 &
  \textbf{2} \\
Datahub &
  3 &
  3 &
  3 &
  3 &
  3.5 &
  3.5 &
  3 &
  3 &
  3 &
  \textbf{3} &
  3 &
  3 &
  3 &
  \textbf{3} &
  3 &
  \textbf{3} &
  3 &
  3 &
  3 &
  4 &
  3 &
  3 &
  2 &
  3 &
  \textbf{3} &
  \textbf{3} &
  3 \\ \bottomrule
\end{tabular}%
}
\caption{\label{ratingstable} Median Difficulties per \KR\ per Question Using Likert Scale(1-5): * $p < 0.05$ for one \KR, ** $p < 0.05$ for both \KR s}
\end{table*}

%% file: sections/relanddisc.tex
\section{Existing Efforts \& Related Work}
\label{sec:relanddisc}
We discuss how our \KR\ complements existing metadata management efforts (Section \ref{sec:discussion}) and related work (Section \ref{sec:relatedwork}).
\input{sections/discussion.tex}
\input{sections/relatedwork.tex}

%% file: sections/discussion.tex
\subsection{Connection to Metadata Efforts}
\label{sec:discussion}
We discuss how the 5W1H+R \KR\ complements existing metadata management initiatives here:

\mypar{FAIR Principles} Catalogs are designed to
manage metadata, which is a keystone to making datasets findable, accessible,
interoperable, and reusable.
A comprehensive, comprehensible, easy-to-use \KR\ for representing metadata can help provide data users with a common framework for understanding a dataset, facilitating reusability.

\mypar{Ontologies} Although we have argued against ontologies
as a catalog's \KR , they can annotate contents within a MI, hence
complementing \KR s by providing more specific terminology for MIs within a partition. 
Concretely, annotations describe
the meaning of MI's keys and values. For example, one can use W3C
DCAT's terms of "dataset distribution", and "temporal
coverage" as What-profile key names, since these can be found by looking at the
data. 

\mypar{ML model tracking} MLFlow~\cite{mlflow_site} and
ModelDB~\cite{modeldb_site} track the lifecycle of machine learning model
engineering and deployment. Both systems produce metadata that describes
training datasets and models. Such metadata can be represented in the 5W1H+R
\KR. First, models and datasets are data assets in the \KR. The metadata
generated can be mapped to partitions in the \KR . For example, the location of
the code file, model parameters, and metrics become How-profiles. A provenance
Relationship captures the link between model and the data used to fit the model,
as well as plots and other derived data products. Model annotations and
descriptions can be represented using the What-profile table of a 5W1H+R data
catalog. Finally, different runs fit well with the versioning of What-Profiles
along with the use of When-Profiles.

\mypar{Feature Stores} Many feature stores have been created to store features \cite{feastfs, sagemfs, tectonfs, hopfs}, along with all metadata required to choose features to build and train a model.
However, users either store feature metadata in whichever format they wish, or in a way that is directly tied to their current ML workflow~\cite{tectonfs}. 
A 5W1H+R data catalog can complement feature stores by allowing users to provide more comprehensive metadata about each feature. This would help users make a more holistic determination as to whether the feature should be included in their training data, and provide other users a rough understanding of the metadata provided for a feature.


\mypar{Metadata Collection} Methods and standards for metadata
collection include Montreal Data Licenses \cite{montreallicense}, Google
Model Cards \cite{googlemodelcards}, Datasheets for Datasets
\cite{gebru2018datasheets}, CancerGrid \cite{cancergrid}, DLHub \cite{dlhub},
and ISO/IEC 11179 Metadata Registry \cite{isoregistry}.

A 5W1H+R data catalog complements these efforts. Metadata collected following
those standards fits well into the 5W1H+R. For example, the Montreal Data
License information could be stored in an asset's Who-Profile and Why-profiles,
as answers to who is allowed to access the data asset, and for what purposes.
Datasheets~\cite{gebru2018datasheets} and other metadata representation formats
fit directly into the \KR .

\mypar{Metrics Layers} A metrics layer is a central repository for code to compute common metrics different users may need (i.e. KPIs)\cite{metriclayer}. This concept is subsumed by the How-profile definition, but may be a useful category to include within the How-profile. 
In the How-profile, one would include not only metrics, such as KPIs, but also other types of usage, such as data preprocessing code. Making a separation between commonly computed metrics and other types of metadata about data usage may prove useful.

%% file: sections/relatedwork.tex
\subsection{Related Work}
\label{sec:relatedwork}

\mypar{Data Reuse Across Data Science Teams} Several studies concerning
collaborations in data science have observed the need for data reuse across data
science teams in an organization, but a lack of solutions that allow for its
reuse \cite{dsinswteams, trustinds, zhangdscollabwork, mullerdscapture,
hilltrials}.  We partially tackle this larger problem by providing a mental
model for metadata (information about data) which allows for information about
data to be represented in a way that can be understood by a large and varied set
of users.  A comprehensive, comprehensible \KR\ can help facilitate data reuse
among data scientists by making it easier for them to develop a shared
understanding about the metadata they have for a data asset.

\mypar{Existing Metadata Definitions} Metadata definitions aim to clarify what
metadata means. One approach to defining metadata is to differentiate 'business'
from 'technical' metadata~\cite{okeratechvsbus, googletechvsbus}.  Ground
proposes defining metadata as Application, Behavior,
Context~\cite{hellerstein2017ground}. However, we do not regard many of these
metadata definitions as \KR s, because they are generally used to explain the
features of a catalog rather than be implemented for metadata storage in a
catalog. Although Ground \cite{hellerstein2017ground} implements its metadata
definition for metadata storage, we do not consider it a \KR\ because Ground's
ABC's do not describe a method to partition the landscape of metadata that is
useful for data management tasks. Instead, it focuses only on
describing the functions of metadata in the data lifecycle.

\mypar{Existing Metadata Models} Metadata models not only define metadata, but also implement that definition. While all models emphasize comprehensiveness, they each apply to different settings and solve various problems. HANDLE~\cite{handlemodel} is a metadata model intended to be generic (similar to comprehensiveness) and respect data lake zones, which facilitate access control.
MEDAL~\cite{medalmodel} defines metadata with respect to several use cases including semantic enrichment, usage tracking, versioning, etc. M4DG~\cite{m4dgmodel} defines metadata with respect to describing data goods in a marketplace.

While the 5W1H+R \KR\ can be considered a metadata model, the 5W1H+R focuses on comprehensibility: providing a shared understanding of data among diverse sets of users. These works may implicitly assume comprehensibility, but do not formally define or evaluate it.

\mypar{Cognitive Fit Theory} The theory of cognitive fit originates in
information systems (IS) \cite{cognitivefitDef}, where it has been used to
inform conceptual schemas~\cite{cog_fitschema1, cog_fitschema2}, such as ER
diagrams. To our knowledge the 5W1H+R \KR\ is the first to use
this theory to propose a \KR\ for metadata management.

\noindent\textbf{Metadata Extraction} is complementary to storage.  Landmark grammars have been used to more effectively solve web data extraction problems~\cite{smallestmeta}. Glean is used to extract structured information from templatic documents~\cite{gleanext}. \cite{prefqueryext} extracts metadata about taxonomical data to answer logic-based preference queries.
Aurum ingests and
processes structured data to discover relationships between similar datasets,
which it models as an enterprise knowledge graph \cite{fernandez2018aurum}.
Juneau \cite{zhang2020finding} discovers related tables using workflows
provided by Jupyter notebooks. Pytheas \cite{pytheas} discovers tables from CSV
files using a flexible rule set. Survey work~\cite{dataprofiling} has explained
some useful dataset profiles, which we call metadata.

\noindent\textbf{Data Provenance} Tuple provenance uses terminology
similar to the 5W1H+R MM, why-, how-, and where-provenance, even though the
meaning is different~\cite{5W1Hprov}. In the ontology space, the
PROV-O ontology for data provenance~\cite{provoont} also uses similar
terminology although for a different purpose. These concepts and
applications are orthogonal to those discussed in this work.

\mypar{Comparisons of Catalog Implementations} \update{R1D1, R2D1-1}{Reports compare existing data catalog implementations using metrics such as
UIs and machine learning} (\ie \update{}{how many tasks are automated) such as
the Forrester Wave}\cite{forresterwave} and the Gartner Magic Quadrant
\cite{gartnerquadrant}\update{}{. We do not compare catalog implementations in
this work.}

\mypar{Other High-Level Modeling Frameworks} Microsoft
Repository\cite{microsoftrepo} provides a model for storing data that is
intended to reduce the number of new terms that the large and varied base of
ActiveX users must learn to store data, while also being extensible, so users
can define their own methods to store data in the way they think is best. We see
parallels between these goals and the ones we propose in this paper: a
comprehensive, comprehensible \KR\ is a model that reduces the number of terms
catalog users must learn to store metadata, while remaining "extensible" in the
sense that users can define metadata however they wish within its partitions.

\mypar{Other 5W1H-based Frameworks} \suggestion{R1O4}{5W1H-based frameworks have been proposed for describing various scientific and computational scenarios, such as describing user events that should trigger applications}~\cite{jang20055w1h}, \suggestion{}{providing a complete description of next-generation probiotics}~\cite{probiotics5w1h}, \suggestion{}{generating information threads and identifying temporal relationships}~\cite{infthread5w1h}, \suggestion{}{and creating public service ontologies}~\cite{pubserv5w1h}.
\suggestion{}{However, these works are similar to the 5W1H+R MM proposed in this paper only at a high level. This work proposes the novel concept of using a 5W1H-based \emph{mental model for metadata} that can be implemented in \emph{data catalogs} to store metadata.}

%% file: sections/conclusion.tex
\section{Conclusion}
\label{sec:conclusions}

As data management percolates more organizations and reaches more
people than ever, many human problems are surfacing. Metadata management is a
critical task to effectively extract value from data. In this paper, we
identified that data catalogs without a well-defined mental model do not
generate agreement among data users, who must collaborate (in storing and
retrieving metadata) to complete their tasks. We then proposed
a new \KR\ for metadata management, informed by an in-depth
study of existing catalog technology. We justified the use of the \KR\ based on
the cognitive fit theory. We evaluated the new \KR\ with a user study. All in all, we consider our work to help shed some light in the vast area of metadata
management.